\documentclass[twocolumn,showpacs,preprintnumbers,amsmath,amssymb]{revtex4}

\usepackage{graphicx}
\usepackage{dcolumn}
\usepackage{bm}

\begin{document}

\title{The Theoretical Proof for GLHUA EM Invisible Double Layer Cloak 
\\
 By Using GL No Scattering Modeling and Inversion
 }

\author{Jianhua Li}
 \altaffiliation[Also at ]{GL Geophysical Laboratory, USA, glhua@glgeo.com}
\author{  Feng Xie, Lee Xie, Ganquan Xie}%
\affiliation{%
GL Geophysical Laboratory, USA
}%

\hfill\break
\author{Ganquan Xie}
\affiliation{
Chinese Dayuling Supercomputational Sciences Center, China
}%

\date{December 30,2016}

\begin{abstract}
In December of 2016, we submitted 3 papers to arXiv for GLHUA double layer cloak with relative parameters not less than 1. The first paper arXiv:1612.02857,arXiv:1701.00534, and arXiv:1701.02583 have already been published in arXiv. 

We proposed the GLHUA double layer cloak; proved the properties of
GLHUA double layer cloak; Using GL no scattering inversion and the pre cloak condition 6.1 to 6.4 in paper [1], we create GLHUA outer layer
cloak radial relative parameter and angular relative parameter theoretically. We proved theorem 4.1 to
theorem 4.4 that the phase velocity of the electromagnetic wave in GLHUA outer layer
cloak is less than light speed and tends to zero at the boundary $r=R_1$, the EM
wave tends to zero at $r=R1$ in regularizing; When source  $r_s > R_2$ in the outside of the cloak and observer $r_o < R_o$ in the concealment, then $\vec E(\vec r_o)=0$ and
$\vec H(\vec r_o)=0$ that is proved in theorem 5.1 and 5.3. We prove that the
EM wave excited in outside of the cloak can not propagation penetrate into the
concealment. In theorem 5.4 to 5.6, we rigorously proved the EM wave excited in outside of the cloak can not be disturbed by the cloak. In theorem 6.1 to theorem 6.6, we prove that the EM wave excited in the concealment can not propagate to outside of GLHUA inner cloak. we prove that the EM wave excited in the concealment can not
be disturbed by the cloak. We theoretically prove that the GLHUA double cloak
is invisible cloak with concealment and with relative parameter not less than 1;
the GLHUA double layer cloak is practicable.
Recently, we discovered a new version of GLHUA outer layer cloak. A GLHUA expansion method to find an exact analytical EM wave propagation in the GLHUA double cloak is detailed proposed in arXiv:1706.10147[6]. Our analytical EM wave in GLHUA cloak is undisputed evidence and rigorous proof to prove that the GLHUA double layer cloak is practicable invisible cloak without exceed light speed propagation. That introduces reader to understand our GLHUA invisible cloak. Our analytical EM wave in GLHUA cloak is undisputed evidence and rigorous proof to prove that  GLHUA double layer invisible cloak and their theoretical proof in  are right.   We completely solved invisible cloak problem in the spherical annular devices. The idea and ingredients in this paper are breakthrough progress and different from all other cloak in other research publications.  Patent of the GLHUA EM cloaks,GLHUA sphere and GL modeling
and inversion
methods are reserved by authors in GL Geophysical Laboratory.
\end{abstract}

\maketitle

\section{\label{sec:level1}INTRODUCTION} 
In paper [1],we create GLHUA sphere $r < R_2$. and proved GLHUA sphere is invisible sphere; the phase velocity in GLHUA sphere is less than light speed and tends to zero at origin, and the electromagnetic wave tends to zero in regularizing at origin.  In this paper, our research working is focus on the 
GLHUA outer layer cloak $R_1 \le r \le R_2$ and GLHUA inner layer cloak,
$R_0 \le r \le R_1$. The research task and content in this paper are different from paper [1] and paper [2] ( paper arXiv:1612.02857). This paper and paper [1] are theoretical base and proof of the paper arXiv:1612.02857 [2].     
The main task and content of this paper is organized as follows: The introduction is presented in section 1. In section 2, we propose the relative electric permittivity and magnetic permeability of GLHUA outer layer cloak in (1) in subsection 2.A; In theorem 2.1, we prove that the EM relative parameters and their derivative are continuous functions in the outer layer cloak and continuous across the outer boundary $r = R_2$. The EM relative parameters in GLHUA outer layer cloak are not less than 1 that is proved in the theorem 2.2; The theorem 2.1 and 2.2 are presented in subsection 2.B;
In subsection 2.C, the relative parameters of the GLHUA inner layer cloak is proposed in (12); In theorem 2.3, we prove that the EM relative parameters and their derivative are continuous functions in the inner layer cloak and continuous across the inner boundary $r = R_0$;
The EM relative parameters in GLHUA inner layer cloak (12) are not less than 1 that is proved in the theorem 2.4; The theorem 2.3 and 2.4 are presented in subsection 2.D. In section 3, we create the relative EM parameters of the GLHUA outer layer cloak; In subsection 3.A, we create relative radial electric permittivity $\varepsilon _{p,r} (r) = 1$ and radial relative magnetic permeability, $\mu _{p,r} (r) = 1$ in GLHUA outer layer cloak; By GL no scattering inversion, we create the inverse of the induced radial parameter by a pending transform to be the radial parameter in GLHUA sphere in [1], the GLHUA radial parameters satisfy the pre cloak condition (6.1) to (6.3) in [1]; To create angular relative electric permittivity and magnetic permeability in physical  GLHUA outer layer cloak that is proposed in subsection 3.B; The pre cloak condition 6.4 in paper [1] is key important to create angular relative electric permittivity and magnetic permeability  in the GLHUA outer layer cloak; From the condition 6.4,
we find the novel GL transform $r_p  = R_1  + Ae^{ - \frac{B}{r}}$
and created the relative parameters of GLHUA outer layer cloak. In section 4, we prove that the phase velocity of electromagnetic wave propagation is less than light speed and tends to zero at the inner boundary, $r=R_1$; The EM wave in outer layer can not 
be arrived to inner boundary $r=R_1$; In theorem 4.1, the $ \mathop {\lim }\limits_{r_p  \to R_1 } E_{p,r} (\vec r_p ) = 0,
 $ and $ \mathop {\lim }\limits_{r_p  \to R_1 } H_{p,r} (\vec r_p ) = 0,
 $ are proved; In theorem 4.2, $ \mathop {\lim }\limits_{r_p  \to R_1 } \frac{1}{{\varepsilon _{p,\theta } }}\frac{\partial }{{\partial r_p }}E_{p,r} (\vec r_p ) = 0,$ and $ \mathop {\lim }\limits_{r_p  \to R_1 } \frac{1}{{\mu _{p,\theta } }}\frac{\partial }{{\partial r_p }}H_{p,r} (\vec r_p ) = 0,$ 
are proved; The  angular EM wave field and their derivative in the GLHUA outer propagation are going to zero in regularizing at inner boundary $r = R_1$ that are  proved in theorem 4.3 and theorem 4.4. In theorem 5.1 to theorem 5.3 in section 5, we proved that
suppose that in the outer sphere annular layer of GLHUA double layer cloak, 
$R_1  < r \le R_2 $, the anisotropic relative electric permittivity and magnetic permeability parameter is in (1); in the inner sphere annular layer GLHUA double cloak,$R_0  < r \le R_1 $, the relative anisotropic electric permittivity and magnetic permeability is in  (12), the source $r_s  > R_2 $  and observer $r_0 < R_1$, then $ \vec E(\vec r_o ) = 0,
$ and $ \vec H(\vec r_o ) = 0$; In the theorem 5.1 to 5.3, we rigorously prove that the incident electromagnetic wave excited outside GLHUA cloak can not propagation penetrate into the sphere $r < R_1$ and can not penetrate into the concealment. In theorem 5.4 to theorem 5.6 in the section 5, we prove that suppose that in the outer sphere annular layer of GLHUA double layer cloak, $R_1  < r \le R_2 $, the anisotropic relative electric permittivity and magnetic permeability parameter is proposed in (1); in the inner sphere annular layer GLHUA double cloak,$R_0  < r \le R_1 $, the relative anisotropic electric permittivity and magnetic permeability is proposed in (12), the source $r_s > R_2$ and observer $r_o > R_2$, then $ \vec E(\vec r_o) = \vec E^b (\vec r_o)$ and $ \vec H(\vec r_o) = \vec H^b (\vec r_o) $.  In the theorem 5.4 to 5.6, we prove
that the incident electromagnetic wave excited outside GLHUA cloak can not be disturbed by the cloak. We theoretically proved the GLHUA outer layer cloak is invisible cloak. Moreover, we theoretically proved the GLHUA inner layer cloak
is invisible cloak. Summary, we theoretically proved the GLHUA double layer cloak
is invisible cloak with concealment. The relative electromagnetic parameters in
GLHUA double layer cloak are not less that 1 that makes that the electromagnetic
wave propagation in GLHUA double layer cloak without infinite speed and without
exceeding light speed; the reciprocal principle is satisfied in GLHUA double layer cloak. The GLHUA double layer cloak is practicable. In the section 7,
GL electromagnetic Eikonal equation for anisotropic material in GLHUA cloak
is proposed. From GL EM Eikonal equation, in GLHUA outer layer cloak,
 wave front is discontinuous and splitting; the ray propagation is complicated and discontinuous at boundary $r=R_1$.  The dicscussion
and conclusion are presented in section 8.

\section {GLHUA double layer electromagnetic cloak with relative parameter not less than 1 }

In this section, we prove that the relative parameters of GLHUA EM double layer
cloak are not less than 1; the relative parameters and their derivatives are continuous in the double layer domain; the relative parameters and their derivatives in outer layer cloak are continuous across the outer boundary $r=R_2$; the relative parameters and their derivatives in inner layer cloak are continuous across the inner boundary $r=R_0$.

\subsection{The relative parameters of the GLHUA outer layer cloak}
we propose GLHUA outer layer cloak in GLHUA double cloak. In the outer sphere annular layer $R_1  < r \le R_2 $, the relative anisotropic electric permittivity and magnetic permeability in Maxwell equation ((1)-(4) in [1]) are

\begin{equation}
\begin{array}{l}
 \varepsilon _r  = \mu _r  = 1, \\ 
 \varepsilon _\theta   = \varepsilon _\phi   = \mu _\theta   = \mu _\phi   =  \\ 
  = \frac{1}{2}\left( {\left( {\frac{{r - R_1 }}{{R_2  - R_1 }}} \right)^\alpha   + \left( {\frac{{R_2  - R_1 }}{{r - R_1 }}} \right)^\alpha  } \right), \\ 
 R_1  < r \le R_2 , \\ 
 0 < \alpha _0  < \alpha  < \alpha _1  < 2, \\ 
 \end{array}
\end{equation}

\subsection{ Properties of GLHUA outer Layer Electromagnetic Cloak}
${\boldsymbol{Theorem \ 2.1: }}$ \ 
In outer layer of GLHUA double layer cloak, the relative electric permittivity and magnetic permeability parameter in (1) and their derivative are continuous across boundary $r=R_2$, outer boundary of outer annular layer GLHUA double cloak.
\hfill\break

${\boldsymbol{Proof: }}$\ 

Because $\varepsilon _r  = \mu _r  = 1$, the radial relative parameter and its derivative are continuous and
continuous across outer boundary $r=R_2$.
From the angular relative parameters in (1),

\begin{equation}
\begin{array}{l}
 \mathop {\lim }\limits_{r \to R_2 } \varepsilon _\theta  \left( r \right) = \mathop {\lim }\limits_{r \to R_2 } \varepsilon _\phi  \left( r \right) =  \\ 
  = \mathop {\lim }\limits_{r \to R_2 } \mu _\theta  \left( r \right) = \mathop {\lim }\limits_{r \to R_2 } \mu _\phi  \left( r \right) =  \\ 
  = \mu _\theta  \left( {R_2 } \right) = \mu _\phi  \left( {R_2 } \right) =  \\ 
  = \varepsilon _\theta  \left( {R_2 } \right) = \varepsilon _\phi  \left( {R_2 } \right) \\ 
  = \frac{1}{2}\left( {\left( {\frac{{R_2  - R_1 }}{{R_2  - R_1 }}} \right)^\alpha   + \left( {\frac{{R_2  - R_1 }}{{R_2  - R_1 }}} \right)^\alpha  } \right) = 1, \\ 
 \end{array}
\end{equation}
Therefore the relative parameters are continuous across the outer boundary$r=R_2$.
Because
\begin{equation}
\begin{array}{l}
 \frac{d}{{dr}}\varepsilon _\theta   = \frac{d}{{dr}}\varepsilon _\phi   =  \\ 
  = \frac{d}{{dr}}\mu _\theta  (r) = \frac{d}{{dr}}\mu _\phi  (r) \\ 
  = \frac{1}{2}\alpha \left( {\left( {\frac{{r - R_1 }}{{R_2  - R_1 }}} \right)^{\alpha  - 1} \frac{1}{{R_2  - R_1 }} - } \right. \\ 
  - \left( {\frac{{R_2  - R_1 }}{{r - R_1 }}} \right)^{\alpha  - 1} \left. {\frac{{R_2  - R_1 }}{{(r - R_1 )^2 }}} \right), \\ 
 R_1  < r \le R_2 , \\ 
 \end{array}
\end{equation}
When $r=R_2$,
\begin{equation}
\begin{array}{l}
 \left. {\frac{d}{{dr}}\mu _r (r)} \right|_{r = R_2 }  \\ 
  = \left. {\frac{d}{{dr}}\varepsilon _r } \right|_{r = R_2 }  = 0, \\ 
 \end{array}
\end{equation}
\begin{equation}
\begin{array}{l}
 \left. {\frac{d}{{dr}}\varepsilon _\theta  } \right|_{r = R_2 }  = \left. {\frac{d}{{dr}}\varepsilon _\phi  } \right|_{r = R_2 }  \\ 
  = \left. {\frac{d}{{dr}}\mu _\theta  (r)} \right|_{r = R_2 }  = \left. {\frac{d}{{dr}}\mu _\phi  (r)} \right|_{r = R_2 }  \\ 
  = \alpha \left( {\frac{1}{{R_2  - R_1 }} - \frac{1}{{R_2  - R_1 }}} \right) = 0, \\ 
 \end{array}
\end{equation}
Therefore, the derivatives of the relative parameters are continuous across the outer boundary $r=R_2$.
  Theorem 2.1 is proved.
\hfill\break

${\boldsymbol{Theorem \ 2.2: }}$ \ 
The relative electric permittivity and magnetic permeability parameters in GLHUA outer layer cloak are not less than 1,  
\begin{equation}
\begin{array}{l}
 \varepsilon _r  = \mu _r  \ge 1, \\ 
 \varepsilon _\theta   = \varepsilon _\phi   = \mu _\theta   = \mu _\phi   \ge 1, \\ 
 \end{array}      
\end{equation}          
      
\hfill\break
${\boldsymbol{Proof: }}$ \ 
For $\alpha  \ne 1$�� the second order derivative of the outer layer parameter (1) is

\begin{equation}
\begin{array}{l}
 \frac{{d^2 }}{{dr^2 }}\varepsilon _\theta   = \frac{{d^2 }}{{dr^2 }}\varepsilon _\phi   \\ 
  = \frac{{d^2 }}{{dr^2 }}\mu _\theta  (r) = \frac{{d^2 }}{{dr^2 }}\mu _\phi  (r) \\ 
  = \frac{1}{2}\alpha \frac{d}{{dr}}\left( {\left( {\frac{{r - R_1 }}{{R_2  - R_1 }}} \right)^{\alpha  - 1} \frac{1}{{R_2  - R_1 }}} \right. \\ 
  - \left. {\frac{{(R_2  - R_1 )^\alpha  }}{{(r - R_1 )^{\alpha  + 1} }}} \right) \\ 
  = \frac{1}{2}\alpha \left( {(\alpha  - 1)\left( {\frac{{r - R_1 }}{{R_2  - R_1 }}} \right)^{\alpha  - 2} } \right. \\ 
 \left. {\left( {\frac{1}{{R_2  - R_1 }}} \right)^2  + (\alpha  + 1)\frac{{\left( {R_2  - R_1 } \right)^\alpha  }}{{\left( {r - R_1 } \right)^{\alpha  + 2} }}} \right), \\ 
 R_1  < r \le R_2 , \\ 
 \end{array}
\end{equation}
\begin{equation}
\begin{array}{l}
 \frac{{d^2 }}{{dr^2 }}\varepsilon _\theta   = \frac{{d^2 }}{{dr^2 }}\varepsilon _\phi   =  \\ 
  = \frac{{d^2 }}{{dr^2 }}\mu _\theta  (r) = \frac{{d^2 }}{{dr^2 }}\mu _\phi  (r) \\ 
  = \frac{1}{2}\alpha \frac{d}{{dr}}\left( {\left( {\frac{{r - R_1 }}{{R_2  - R_1 }}} \right)^{\alpha  - 1} } \right. \\ 
 \left. {\frac{1}{{R_2  - R_1 }} - \frac{{(R_2  - R_1 )^\alpha  }}{{(r - R_1 )^{\alpha  + 1} }}} \right) \\ 
  = \frac{1}{2}\alpha (\alpha  - 1)\left( {\frac{{r - R_1 }}{{R_2  - R_1 }}} \right)^{\alpha  - 2}  \\ 
 \left( {\frac{1}{{R_2  - R_1 }}} \right)^2 \left( {1 - \frac{{(\alpha  + 1)}}{{1 - \alpha }}\frac{{\left( {R_2  - R_1 } \right)^{2\alpha } }}{{\left( {r - R_1 } \right)^{2\alpha } }}} \right) \\ 
  \ge 0,R_1  < r \le R_2 , \\ 
 \end{array}
\end{equation}
For  $\alpha  = 1$ the second order derivative of the outer layer parameter (1) is

\begin{equation}
\begin{array}{l}
 \frac{{d^2 }}{{dr^2 }}\varepsilon _\theta   = \frac{{d^2 }}{{dr^2 }}\varepsilon _\phi   =  \\ 
  = \frac{{d^2 }}{{dr^2 }}\mu _\theta  (r) = \frac{{d^2 }}{{dr^2 }}\mu _\phi  (r) \\ 
  = \frac{1}{2}\alpha \frac{d}{{dr}}\left( {\frac{1}{{R_2  - R_1 }} - \frac{{(R_2  - R_1 )}}{{(r - R_1 )^2 }}} \right) \\ 
  = \frac{1}{2}\left( {2\frac{{\left( {R_2  - R_1 } \right)}}{{\left( {r - R_1 } \right)^3 }}} \right) \ge 0,R_1  < r \le R_2 , \\ 
 \end{array}
\end{equation}
 The first order derivative function of GLHUA outer layer cloak parameter (1)
   \[
\begin{array}{l}
 \frac{d}{{dr}}\mu _\theta  (r) = \frac{d}{{dr}}\mu _\phi  (r) \\ 
  = \frac{d}{{dr}}\varepsilon _\theta   = \frac{d}{{dr}}\varepsilon _\phi  , \\ 
 \end{array}
\]      
 is monotone increasing function,

and
\begin{equation}
\begin{array}{l}
 \frac{d}{{dr}}\mu _\theta  (r) = \frac{d}{{dr}}\mu _\phi  (r) \\ 
  = \frac{d}{{dr}}\varepsilon _\theta   = \frac{d}{{dr}}\varepsilon _\phi   \le 0, \\ 
 \end{array}
\end{equation}
GLHUA outer layer cloak parameter function (1) is monotone decreasing in interval $[R_1 ,R_2 ]$

\begin{equation}
\begin{array}{l}
 \varepsilon _\theta  (r) = \varepsilon _\phi  (r) =  \\ 
  = \mu _\theta  (r) = \mu _\phi  (r) \\ 
  \ge \varepsilon _\theta  (R_2 ) = 1, \\ 
 \end{array}
\end{equation}
Therefore, the relative parameters in (1) of GLHUA outer layer cloaking material are not less than 1.

\subsection{The relative parameters of the GLHUA inner layer cloak}
We propose relative parameters of the inner cloak in GLHUA double layer cloak and prove that the relative parameters are not less than 1. 
In GLHUA inner annular layer cloak $R_0  < r \le R_1 $, the relative anisotropic electric permittivity and magnetic permeability in Maxwell equation ((1)-(4) in [1]) are

\begin{equation}
\begin{array}{l}
 \varepsilon _r  = \mu _r  = 1, \\ 
 \varepsilon _\theta   = \varepsilon _\phi   = \mu _\theta   = \mu _\phi   =  \\ 
  = \frac{1}{2}\left( {\left( {\frac{{R_1  - r}}{{R_1  - R_0 }}\frac{{R_0 }}{r}} \right)^\alpha   + \left( {\frac{{R_1  - R_0 }}{{R_1  - r}}\frac{r}{{R_0 }}} \right)^\alpha  } \right), \\ 
 R_0  < r \le R_1 , \\ 
 0 < \alpha _0  < \alpha  < \alpha _1  < 2, \\ 
 \end{array}
\end{equation}
\subsection{ Properties of GLHUA inner Layer Electromagnetic Cloak}
${\boldsymbol{Theorem \ 2.3: }}$ \ 
In inner layer of GLHUA double layer cloak, the relative electric permittivity and magnetic permeability parameter in (12) and their derivative are continuous across boundary $r=R_0$, inner boundary of inner annular layer GLHUA double cloak.
\hfill\break

${\boldsymbol{Proof: }}$ \ 

Because $\varepsilon _r  = \mu _r  = 1$, the radial relative parameter and its derivative are continuous and
continuous across inner boundary $r=R_0$.
From the angular relative parameters in (12),

\begin{equation}
\begin{array}{l}
 \mathop {\lim }\limits_{r \to R_0 } \varepsilon _\theta  \left( r \right) = \mathop {\lim }\limits_{r \to R_0 } \varepsilon _\phi  \left( r \right) =  \\ 
  = \mathop {\lim }\limits_{r \to R_0 } \mu _\theta  \left( r \right) = \mathop {\lim }\limits_{r \to R_0 } \mu _\phi  \left( r \right) =  \\ 
  = \mu _\theta  \left( {R_0 } \right) = \mu _\phi  \left( {R_0 } \right) =  \\ 
  = \varepsilon _\theta  \left( {R_0 } \right) = \varepsilon _\phi  \left( {R_0 } \right) \\ 
  = \frac{1}{2}\left( {\left( {\frac{{R_1  - R_0 }}{{R_1  - R_0 }}\frac{{R_0 }}{{R_0 }}} \right)^\alpha   + \left( {\frac{{R_1  - R_0 }}{{R_1  - R_0 }}\frac{{R_0 }}{{R_0 }}} \right)^\alpha  } \right) = 1, \\ 
 \end{array}
\end{equation}
Therefore the the radial and angular relative parameters in (12) of the inner cloak are continuous in domain and continuous across inner boundary $r=R_0$.
Because
\begin{equation}
\begin{array}{l}
 \frac{d}{{dr}}\varepsilon _\theta   = \frac{d}{{dr}}\varepsilon _\phi   =  \\ 
  = \frac{d}{{dr}}\mu _\theta  (r) = \frac{d}{{dr}}\mu _\phi  (r) \\ 
  = \frac{1}{2}\alpha \left( {\left( {\frac{{R_1  - r}}{{R_1  - R_0 }}\frac{{R_0 }}{r}} \right)^{\alpha  - 1}  + \left( {\frac{{R_1  - R_0 }}{{R_1  - r}}\frac{r}{{R_0 }}} \right)^{\alpha  - 1} } \right) \\ 
 \left( {\frac{{R_1  - R_0 }}{{(R_1  - r)^2 }}\frac{r}{{R_0 }} - \frac{1}{{R_1  - R_0 }}\frac{{R_0 }}{r} + } \right. \\ 
 \left. { + \frac{{R_1  - R_0 }}{{(R_1  - r)}}\frac{1}{{R_0 }} - \frac{{R_1  - r}}{{R_1  - R_0 }}\frac{{R_0 }}{{r^2 }}} \right), \\ 
 R_0  < r \le R_1 , \\ 
 \end{array}
\end{equation}
when in the inner boundary $r=R_0$,

\begin{equation}
\begin{array}{l}
 \left. {\frac{d}{{dr}}\mu _r (r)} \right|_{r = R_0 }  = \left. {\frac{d}{{dr}}\varepsilon _r } \right|_{r = R_0 }  = 0, \\ 
 \left. {\frac{d}{{dr}}\mu _\theta  (r)} \right|_{r = R_0 }  = \left. {\frac{d}{{dr}}\mu _\phi  (r)} \right|_{r = R_0 }  \\ 
  = \left. {\frac{d}{{dr}}\varepsilon _\theta  } \right|_{r = R_0 }  = \left. {\frac{d}{{dr}}\varepsilon _\phi  } \right|_{r = R_0 }  = 0, \\ 
 \end{array}
\end{equation}
Therefore the the radial and angular relative parameter in (12) of the inner cloak and its derivative are continuous in domain and continuous across inner boundary $r=R_0$.Theorem 2.3 is proved.
\hfill\break

${\boldsymbol{Theorem \ 2.4: }}$ \ 
The relative electric permittivity and magnetic permeability parameters in (12) in GLHUA inner layer cloak are not less than 1.
 \begin{equation}
\begin{array}{l}
 \varepsilon _r  = \mu _r  \ge 1, \\ 
 \varepsilon _\theta   = \varepsilon _\phi   = \mu _\theta   = \mu _\phi   \ge 1, \\ 
 \end{array}      
\end{equation} 

\hfill\break
${\boldsymbol{Proof: }}$ \

From GLHUA inner cloak parameters (12), we calculate and re arrange their first order derivative,

\begin{equation}
\begin{array}{l}
 \frac{d}{{dr}}\varepsilon _\theta   = \frac{d}{{dr}}\varepsilon _\phi   =  \\ 
  = \frac{d}{{dr}}\mu _\theta  (r) = \frac{d}{{dr}}\mu _\phi  (r) \\ 
  = \frac{1}{2}\alpha \left( {\left( {\frac{{R_1  - r}}{{R_1  - R_0 }}\frac{{R_0 }}{r}} \right)^{\alpha  - 1} } \right. \\ 
  + \left. {\left( {\frac{{R_1  - R_0 }}{{R_1  - r}}\frac{r}{{R_0 }}} \right)^{\alpha  - 1} } \right) \\ 
 \left( {\frac{{R_1  - R_0 }}{{(R_1  - r)^2 }}\frac{r}{{R_0 }}\left( {1 - \left( {\frac{{(R_1  - r)}}{{R_1  - R_0 }}\frac{{R_0 }}{r}} \right)^2 } \right)} \right. \\ 
 \left. { + \frac{{R_1  - R_0 }}{{(R_1  - r)}}\frac{1}{{R_0 }}\left( {1 - \left( {\frac{{R_1  - r}}{{R_1  - R_0 }}\frac{{R_0 }}{r}} \right)^2 } \right)} \right) \ge 0, \\ 
 R_0  < r \le R_1 , \\ 
 \end{array}
\end{equation}
Then parameter function (12) of the inner layer cloak is monotone increasing in the interval
$[R_0 ,R_1 ]$,
\hfill\break

\begin{equation}
\begin{array}{l}
 \varepsilon _\theta  (r) = \varepsilon _\phi  (r) =  \\ 
  = \mu _\theta  (r) = \mu _\phi  (r) \\ 
  \ge \varepsilon _\theta  (R_0 ) \ge 1, \\ 
 \end{array}
\end{equation}
The relative parameters (12) GLHUA inner layer cloak material  are not less than 1.
Theorem 2.4 is proved.

The theorem 2.2 shows that the relative parameters (1) of GLHUA outer layer cloak is not less than 1,  and  theorem 2.4 shows that the relative parameters (12) of GLHUA inner layer cloak is not less than 1. The electromagnetic wave propagation in GLHUA double cloak without exceeding light speed and without infinite speed propagation.
 
\section{Create GLHUA outer layer cloak} 

In our paper [1], we used $\vec r$ and $\vec r'$ to denote the vector, used $r$ and $r'$ to denote radial
coordinate in GLHUA invisible virtual sphere and in the free space. For avoiding confusion, in this section, We use $r_p$, ${r'}_p$  and  $\vec {r_p}$ to denote radial coordinate and vector in physical space, respectively. The $r$, $\vec r$  in section 2 is same as $r_p$ and $\vec {r_p}.$ in this section.
In this section, we create GLHUA cloak material for outer layer $R_1  \le r_p  \le R_2 $ of GLHUA double layer cloak. 

\subsection{Create relative radial electric permittivity $\varepsilon _{p,r} (r) = 1$ and radial relative magnetic permeability, $\mu _{p,r} (r) = 1$ in GLHUA outer layer cloak}
 
Ideal of creating our GLHUA cloak is as follows. in the first, we create GLHUA cloak relative radial electric permittivity $\varepsilon _{p,r} (r) = 1$  and radial relative magnetic permeability,  $\mu _{p,r} (r) = 1,$ in any thickness physical outer annular layer $R_1  \le r_p  \le R_2 $. There exsits a pending GL transform to map the physical outer annular layer  $R_1  \le r_p  \le R_2 $ to GLHUA virtual sphere $r \le R_2$, the transform induces a relative radial electric permittivity and magnetic permeability. We use the inverse of the transform induced relative radial electric permittivity and magnetic permeability as GLHUA relative permittivity and permeability in the GLHUA virtual invisible sphere  $r \le R_2$, which satisfy GLHUA pre cloak material conditions (6.1)-(6.3) [1] in the invisible virtual sphere  $r \le R_2$.

Suppose that relative permittivity $\varepsilon _p$ and permeability $\mu _p$ in GLHUA outer layer cloak is transformed from GLHUA virtual invisible sphere $r \le R_2$ with relative permittivity $\varepsilon$  and permeability $\mu$ by a novel radial transform which is called GL transform.
 
We propose the GL transform is as follows
\begin{equation}
\begin{array}{l}
 r_p  = R_1  + p(r), \\ 
 r = p^{ - 1} (r_p  - R_1 ), \\ 
 \frac{d}{{dr}}r_p  = \frac{d}{{dr}}p(r), \\ 
 \end{array}
\end{equation}
\begin{equation}
\begin{array}{l}
 \theta _p  = \theta , \\ 
 \phi _p  = \phi , \\ 
 \end{array}
\end{equation}
 
The function $p(r)$ should satisfies the following condition,

\begin{equation}
\begin{array}{l}
 p(0) = 0, \\ 
 p(R_2 ) = R_2  - R_1 , \\ 
 \frac{d}{{dr}}p(R_2 ) = 1, \\ 
 \frac{d}{{dr}}p(r) > 0, \\ 
 \end{array}
\end{equation}

The transform induces a relative radial electric permittivity and magnetic permeability,

\begin{equation}
\begin{array}{l}
 \varepsilon _{p,r} (r) = \frac{{r^2 }}{{r_p ^2 }}\frac{{dp(r)}}{{dr}}\varepsilon _r (r), \\ 
 \mu _{p,r} (r) = \frac{{r^2 }}{{r_p ^2 }}\frac{{dp(r)}}{{dr}}\mu _r (r), \\ 
 \end{array}
\end{equation}
Because we chose physical relative parameter $\varepsilon _{p,r} (r) = \mu _{p,r} (r) = 1$, in pysical outer annular layer  $R_1  \le r_p  \le R_2 $, relative parameters in GLHUA sphere, $r \le R_2$, is mapped to be 

\begin{equation}
\begin{array}{l}
 \varepsilon _r (r) = \mu _r (r) =  \\ 
  = \varepsilon _{p,r} (r)\frac{{r_p ^2 }}{{r^2 }}\frac{{dr}}{{dp(r)}} =  \\ 
  = \frac{{r_p ^2 }}{{r^2 }}\frac{{dr}}{{dp(r)}}. \\ 
 \end{array}
\end{equation}

It is easy to check that the relative parameters $\varepsilon _r (r)$ and $\mu _r (r)$ satisfy GLHUA pre cloak material conditions (6.1)-(6.3) in [1]. Suppose that GL electric wave $E(\vec r) = \varepsilon _r r^2 E_r (\vec r)$ in (18) in [1]  is solution of the GL radial electric wave equation (19) in [1], substitute (23) into (18) in GLHUA virtual sphere in [1], we have

\begin{equation}
\begin{array}{l}
 E(\vec r) = \varepsilon _r r^2 E_r (\vec r) \\ 
  = \frac{{r_p ^2 }}{{r^2 }}\frac{{dr}}{{dp(r)}}r^2 E_r (\vec r) \\ 
  = r_p ^2 \frac{{dr}}{{dp(r)}}E_r (\vec r), \\ 
 \end{array}
\end{equation}
because induced electric wave in annular layer $R_1  \le r_p  \le R_2$ by GL transform is

\begin{equation}
E_{p,r} (\vec r) = \frac{{dr}}{{dp(r)}}E_r (\vec r),
\end{equation}

We have the relationship between induced electric wave field $E_{p,r} (\vec r)$ in annular layer  $R_1  \le r_p  \le R_2$ and GL electric wave field $E(\vec r)$ in virtual sphere $r \le R_2$,

\begin{equation}
\begin{array}{l}
 E_{p,r} (\vec r) = \frac{{dr}}{{dp(r)}}E_r (\vec r) \\ 
  = \frac{1}{{r_p ^2 }}E(\vec r), \\ 
 \end{array}
\end{equation}

and induced magnetic wave is

\begin{equation}
\begin{array}{l}
 H_{p,r} (\vec r) = \frac{{dr}}{{dp(r)}}H_r (\vec r) =  \\ 
  = \frac{1}{{r_p ^2 }}H(\vec r), \\ 
 \end{array}
\end{equation}

where $E(\vec r)$ and  $H(\vec r)$  is GL electromagnetic wave field in GLHUA virtual sphere
$r \le R_2$ in [1].

\subsection{Create angular relative electric permittivity and magnetic permeability in physical  GLHUA outer layer cloak }

In this section, we create angular relative electric permittivity and magnetic permeability, $\varepsilon _{p,\theta } (r_p )$,
$\varepsilon _{p,\phi } (r_p )$,
$\mu _{p,\theta } (r_p )$, 
$\mu _{p,\phi } (r_p )$,  
in physical GLHUA outer layer cloak by GL no scattering inversion,

\begin{equation}
\begin{array}{l}
 \varepsilon _{p,\theta } (r) = \varepsilon _{p,\phi } (r) =  \\ 
  = \varepsilon _\theta  (r)\left( {\frac{{dr_p }}{{dr}}} \right)^{ - 1}  = \varepsilon _\phi  (r)\left( {\frac{{dr_p }}{{dr}}} \right)^{ - 1} , \\ 
 \mu _{p,\theta } (r) = \mu _{p,\phi } (r) =  \\ 
  = \mu _\theta  (r)\left( {\frac{{dr_p }}{{dr}}} \right)^{ - 1}  = \mu _\phi  (r)\left( {\frac{{dr_p }}{{dr}}} \right)^{ - 1} , \\ 
 \end{array}
\end{equation}
In the pre cloak conditions $(6.1) \ to \ (6.4)$ for GLHUA invisible sphere in [1], the condition (6.4) in [1] is key important to create angular relative electric permittivity and relative magnetic permeability in the physical GLHUA outer layer cloak. Recall (6.4) in (59) in [1] 

\begin{equation}
\varepsilon {}_\theta  = f(r)\frac{1}{{r^2 }}, 
\end{equation}

By the relationship formula (28),

\begin{equation}
\varepsilon {}_\theta  = \varepsilon {}_{p,\theta }\frac{{\partial p(r)}}{{\partial r}} = f(r)\frac{1}{{r^2 }},
\end{equation}

If we chose

\begin{equation}
f(r) = \varepsilon {}_{p,\theta }
\end{equation}

by (28), we have
\begin{equation}
\frac{{\partial p(r)}}{{\partial r}} = \frac{1}{{r^2 }},
\end{equation}

\begin{equation}
p(r) =  - \frac{1}{r},
\end{equation}

\begin{equation}
\mathop {\lim }\limits_{r \to 0} p(r) =  - \mathop {\lim }\limits_{r \to 0} \frac{1}{r} =  - \infty,
\end{equation}
that make contradiction with requirement $p(0) = 0$ in (21),
choice (31) is wrong. 
We have to find other right choice that

\begin{equation}
\begin{array}{l}
 \varepsilon {}_\theta  = \varepsilon {}_{p,\theta }\frac{{\partial p(r)}}{{\partial r}} =  \\ 
  = \varepsilon {}_{p,\theta }\frac{{p(r)B}}{{p(r)B}}\frac{{\partial p(r)}}{{\partial r}} =  \\ 
  = f(r)\frac{1}{{r^2 }}, \\ 
 \end{array}
\end{equation}

By GL no scattering inversion, it is best way, we chose

\begin{equation}
f(r) = \varepsilon {}_{p,\theta }p(r)B,
\end{equation}
\hfill\break

and

\begin{equation}
\frac{1}{{Bp(r)}}\frac{{\partial p(r)}}{{\partial r}} = \frac{1}{{r^2 }},
\end{equation}
To solve differential equation (37), we have

\begin{equation}
\int {\frac{{dp}}{{p(r)}}}  = B\int {\frac{{dr}}{{r^2 }}} ,
\end{equation}
\hfill\break

We obtain 
\begin{equation}
p(r) = Ae^{ - \frac{B}{r}} ,
\end{equation}
which satisfies equation (37). Substitute the basic requirement conditions (21) into GL transform (39), the novel GL transform is found
\begin{equation}
\begin{array}{l}
 r_p  = R_1  + p(r) \\ 
  = R_1  + Ae^{ - \frac{B}{r}} , \\ 
 \end{array}
\end{equation}
in the meantime, the pending coefficient $A > 0$, $B > 0$ are determined by the continuous and
derivative continuous condition (21),

\begin{equation}
A = (R_2  - R_1 )e^{\frac{B}{{R_2 }}}, 
\end{equation}

\begin{equation}
B = \frac{{R_2^2 }}{{(R_2  - R_1 )}},
\end{equation}

It is easy to check that the GL transform (40)-(42) satisfy the transform condition (21).

\begin{equation}
f(r) = \varepsilon {}_{p,\theta }p(r)B,
\end{equation}

The GLHUA pre cloak material condition 6.4 in (59) in GLHUA sphere in [1]

\begin{equation}
\mathop {\lim }\limits_{r \to 0} f(r) = \frac{1}{2}R_2 ^2 ,
\end{equation}
and $p(0)=0$ request

\begin{equation}
\varepsilon {}_{p,\theta } = Cp(r) + D\left( {p(r)} \right)^{ - 1},
\end{equation}
Substitute (45) into (43), we have

\begin{equation}
\begin{array}{l}
 f(r) = \varepsilon {}_{p,\theta }p(r)B \\ 
  = \left( {Cp^2 (r) + D} \right)B, \\ 
 \end{array}
\end{equation}
By the GLHUA sphere pre cloak material condition 6.4 in (59) in [1],
\begin{equation}
\mathop {\lim }\limits_{r \to 0} f(r) = \frac{1}{2}R_2 ^2 ,
\end{equation}

\begin{equation}
\begin{array}{l}
 \mathop {\lim }\limits_{r \to 0} f(r) = \mathop {\lim }\limits_{r \to 0} \varepsilon {}_{p,\theta }p(r)B =  \\ 
 \mathop {\lim }\limits_{r \to 0} \left( {Cp^2 (r) + D} \right)B = \frac{{R_2 ^2 }}{2}, \\ 
 \end{array}
\end{equation}

\begin{equation}
D = \frac{{R_2 ^2 }}{{2B}} = \frac{{R{}_2 - R_1 }}{2},
\end{equation}

\begin{equation}
\begin{array}{l}
 \varepsilon {}_{p,\theta }(R_2 ) =  \\ 
  = Cp(R_2 ) + D\left( {p(R_2 )} \right)^{ - 1}  = 1, \\ 
 \end{array}
\end{equation}
By the continuous condition $p(R_2 ) = R_2  - R_1$, we obtain
\begin{equation}
C = \frac{1}{2}\frac{1}{{R_2  - R_1 }},
\end{equation}

Substitute $p(r) = Ae^{ - \frac{B}{r}}  = r_p  - R_1 $ in (40), $C$ in (51), $D$ in (49) into (45), we create GLHUA cloak relative angular material in outer layer $R_1  \le r_p  \le R_2 $,

\begin{equation}
\varepsilon {}_{p,\theta } = \frac{1}{2}\left( {\frac{{r_p  - R_1 }}{{R_2  - R_1 }} + \frac{{R_2  - R_1 }}{{r_p  - R_1 }}} \right),
\end{equation}

It is the relative angular permittivity in GLHUA outer layer cloak, the $\varepsilon {}_{p,\phi }$, $\mu{}_{p,\theta }$, $\mu {}_{p,\phi }$ are same as $\varepsilon {}_{p,\theta }$.  Summary, we complete create the outer annular layer cloak of GLHUA double layer electromagnetic invisible cloak. In outer annular layer $R_1  \le r_p  \le R_2 $, GLHUA cloak relative electric permittivity and magnetic permeability are 
\begin{equation}
\begin{array}{l}
 \varepsilon _{p,r} (r) = \mu _{p,r} (r) = 1, \\ 
 \varepsilon {}_{p,\theta } = \varepsilon {}_{p,\iota } = \mu {}_{p,\theta } = \mu {}_{p,\phi } =  \\ 
  = \frac{1}{2}\left( {\frac{{r_p  - R_1 }}{{R_2  - R_1 }} + \frac{{R_2  - R_1 }}{{r_p  - R_1 }}} \right), \\ 
 \end{array}
\end{equation}
Using above GL no scattering inversion, we theoretically create GLHUA cloak material in outer annular layer $R_1  \le r_p  \le R_2 $, which has been proposed in (1)  with $\alpha  = 1$ for outer layer of GLHUA double layer cloak in the section 2.  For $\alpha  = 1$ , similarly, we can create GLHUA cloak material in outer annular layer $ R_1  \le r_p  \le R_2$  as follows

\begin{equation}
\begin{array}{l}
 \varepsilon _{p,r} (r) = \mu _{p,r} (r) = 1, \\ 
 \varepsilon {}_{p,\theta } = \varepsilon {}_{p,\iota } = \mu {}_{p,\theta } = \mu {}_{p,\phi } =  \\ 
  = \frac{1}{2}\left( {\left( {\frac{{r_p  - R_1 }}{{R_2  - R_1 }}} \right)^\alpha   + \left( {\frac{{R_2  - R_1 }}{{r_p  - R_1 }}} \right)^\alpha  } \right), \\ 
 \end{array}
\end{equation}
which is proposed in (1) in section 2. 
                          
The GL transform (40)-(42) maps outer annular layer $R_1  \le r_p  \le R_2 $ to invisible virtual sphere , maps the GLHUA cloak relative electric permittivity and magnetic permeability (53) in outer layer $R_1  \le r_p  \le R_2 $  to the GLHUA cloak relative radial electric permittivity and magnetic permeability in (23), and relative angular electric permittivity and magnetic permeability in (28) in the virtual invisible sphere space $r \le R_2$,which satisfy GLHUA pre cloak conditions (6.1)-(6.4) in virtual sphere in [1]. By theorem 6.1- 6.4 in[1] , that incident electromagnetic wave excited in outside of virtual sphere $r > R_2$, can not be disturbed by the virtual sphere. The incident electromagnetic wave is smoothly propagating enter the virtual sphere and going to zero, when the r is going to zero, that does derive the following theorem 4.1 to 4.4 of GLHUA cloak theory in outer layer $R_1  \le r_p  \le R_2 $  of GLHUA double layer cloak. 

\section{The electromagnetic field approaching to zero when $r_p$ going to $R_1$ in outer annular layer of GLHUA double layer cloak}
${\boldsymbol{Theorem \ 4.1: }}$\
Suppose that the relative anisotropic relative GLHUA electric permittivity and magnetic permeability
$\varepsilon _{p,r } (r_p )$,
$\mu _{p,r } (r_p )$,
$\varepsilon _{p,\theta } (r_p )$,
$\varepsilon _{p,\phi } (r_p )$,
$\mu _{p,\theta } (r_p )$, 
$\mu _{p,\phi } (r_p )$,   
is denoted by (53), then 
\begin{equation}
\mathop {\lim }\limits_{r_p  \to R_1 } E_{p,r} (\vec r_p ) = 0,
\end{equation}
\begin{equation}
\mathop {\lim }\limits_{r_p  \to R_1 } H_{p,r} (\vec r_p ) = 0,
\end{equation}
${\boldsymbol{Proof: }}$\ 
Because GLHUA cloak radial electric permittivity $\varepsilon _{p,r} (r) = 1$ and magnetic permeability, $\mu _{p,r} (r) = 1$  in physical outer annular layer $R_1  \le r_p  \le R_2,$ , GL transform (40)-(42) one-to-one onto maps outer physical annular layer $R_1  \le r_p  \le R_2,$ to virtual sphere $r\le R_2$�� In the virtual invisible sphere $ r \le R_2$ , GLHUA relative radial electric permittivity and magnetic permeability in (23) is the inverse of the induced relative radial electric permittivity and magnetic permeability by GL transform (40)-(42). It is easy to check the relative permittivity $\varepsilon_r$ and permeability $\mu_r$ in (23) in the virtual sphere satisfy GLHUA pre cloak material conditions (6.1)-(6.3) in [1]. Also, because GLHUA cloak angular electric permittivity and magnetic permeability in physical outer annular layer $R_1  \le r_p  \le R_2,$ is denoted by (53). By GL transform (40)-(42), the corresponding angular electric permittivity and magnetic permeability in virtual invisible sphere$ r \le R_2$ is 
\begin{equation}
\begin{array}{l}
 \varepsilon {}_\theta  = \varepsilon {}_\phi  = \mu {}_\theta  = \mu {}_\phi  \\ 
  = f(r)\frac{1}{{r^2 }} \\ 
  = \frac{1}{2}\left( {1 + e^{2\frac{B}{{R_2 }}} e^{ - 2\frac{B}{r}} } \right)R_2 ^2 \frac{1}{{r^2 }}, \\ 
 \end{array}
\end{equation}
It satisfies GLHUA sphere pre cloak material conditions (6.4) in (59) in [1].
The $\varepsilon _r$ and $\mu _r$  is inverse of radial electric permittivity and magnetic permeability induced by transform (40)-(42) that to be
\begin{equation}
\varepsilon _r  = \mu _r = \frac{{r_p ^2 }}{{r^2 }}\frac{{dr}}{{dr_p }},
\end{equation}
satisfies $(6.1) \ to \ (6.3)$ in [1].
 By applying theorem 6.1, in GLHUA invisible virtual sphere $ r \le R_2 $ in [1], 

\begin{equation}
\mathop {\lim }\limits_{r \to 0} E(\vec r) = 0,
\end{equation}
By (26), the radial electromagnetic wave field in outer physical layer  $R_1  \le r_p  \le R_2,$  induced by the transform (40)-(42) is,
\begin{equation}
\begin{array}{l}
 E_{p,r} (\vec r) = \frac{{dr}}{{dp(r)}}E_r (\vec r) \\ 
  = \frac{{dr}}{{dp(r)}}\frac{1}{{r^2 \varepsilon _r }}E(\vec r) \\ 
  = \frac{{dr}}{{dp(r)}}\frac{{dp(r)}}{{dr}}\frac{{r^2 }}{{r^2 r_p ^2 }} \\ 
  = \frac{1}{{r_p ^2 }}E(\vec r), \\ 
 \end{array}
\end{equation}

\begin{equation}
\begin{array}{l}
 \mathop {\lim }\limits_{r_p  \to R_1 } E_{p,r} (\vec r_p ) = \mathop {\lim }\limits_{r \to 0} E_{p,r} (\vec r) \\ 
  = \mathop {\lim }\limits_{r \to 0} \frac{1}{{r_p (r)^2 }}E(\vec r) \\ 
  = \frac{1}{{R_1 ^2 }}\mathop {\lim }\limits_{r \to 0} E(\vec r) = 0, \\ 
 \end{array}
\end{equation}
The limitation (55) is proved. Similarly, we can prove limitation (56)
\begin{equation}
\mathop {\lim }\limits_{r_p  \to R_1 } H_{p,r} (\vec r_p ) = 0
\end{equation}
Theorem 4.1 is proved.
\hfill\break

${\boldsymbol{Theorem \ 4.2: }}$\
Suppose that the relative anisotropic relative GLHUA electric permittivity and magnetic permeability
$\varepsilon _{p,r } (r_p )$,
$\mu _{p,r } (r_p )$,
$\varepsilon _{p,\theta } (r_p )$,
$\varepsilon _{p,\phi } (r_p )$,
$\mu _{p,\theta } (r_p )$, 
$\mu _{p,\phi } (r_p )$,   
is denoted by (53), then 
\begin{equation}
\mathop {\lim }\limits_{r_p  \to R_1 } \frac{1}{{\varepsilon _{p,\theta } }}\frac{\partial }{{\partial r_p }}E_{p,r} (\vec r_p ) = 0,
\end{equation}
\begin{equation}
\mathop {\lim }\limits_{r_p  \to R_1 } \frac{1}{{\mu _{p,\theta } }}\frac{\partial }{{\partial r_p }}H_{p,r} (\vec r_p ) = 0,
\end{equation}
\hfill\break
${\boldsymbol{Proof: }}$ \ 
By direct calculation, the derivative of the radial electric wave in GLHUA physical outer annular layer cloak, $R_1  \le r_p  \le R_2 $, induced by GL transform (40)-(42) is

\begin{equation}
\begin{array}{l}
 \frac{1}{{\varepsilon _{p,\theta } }}\frac{\partial }{{\partial r_p }}E_{p,r} (\vec r_p ) = \frac{1}{{\varepsilon _{p,\theta } }}\frac{\partial }{{\partial r_p }}E_{p,r} (\vec r_p (r)) \\ 
  = \frac{1}{{\varepsilon _{p,\theta } }}\frac{\partial }{{\partial r_p }}\frac{1}{{r_p ^2 (r)}}E(\vec r) =  \\ 
  = \frac{1}{{\varepsilon _{p,\theta } }}\frac{{\partial r}}{{\partial r_p }}\frac{\partial }{{\partial r}}\frac{1}{{r_p ^2 (r)}}E(\vec r) \\ 
  = \frac{1}{{\varepsilon _{p,\theta } }}\frac{{\partial r}}{{\partial r_p }}\frac{1}{{r_p ^2 (r)}}\frac{\partial }{{\partial r}}E(\vec r) -  \\ 
  - 2\frac{1}{{\varepsilon _{p,\theta } }}\frac{{\partial r}}{{\partial r_p }}\frac{1}{{r_p ^3 (r)}}\frac{{\partial r_p }}{{\partial r}}E(\vec r) \\ 
  = \frac{1}{{r_p ^2 (r)}}\frac{1}{{\varepsilon _\theta  }}\frac{\partial }{{\partial r}}E(\vec r) -  \\ 
  - 2\frac{1}{{\varepsilon _{p,\theta } }}\frac{1}{{r_p ^3 (r)}}E(\vec r), \\ 
 \end{array}
\end{equation}

According to the theorem 6.1 and theorem 6.2 in [1], we have

\begin{equation}
\begin{array}{l}
 \mathop {\lim }\limits_{r_p  \to R_1 } \frac{1}{{\varepsilon _{p,\theta } }}\frac{\partial }{{\partial r_p }}E_{p,r} (\vec r_p ) \\ 
  = \mathop {\lim }\limits_{r \to 0} \frac{1}{{r_p ^2 (r)}}\frac{1}{{\varepsilon _\theta  }}\frac{\partial }{{\partial r}}E(\vec r) -  \\ 
  - \mathop {\lim }\limits_{r \to 0} 2\frac{1}{{\varepsilon _{p,\theta } }}\frac{1}{{r_p ^3 (r)}}E(\vec r) = 0, \\ 
 \end{array}
\end{equation}
(63) is proved. Similarly, we can prove (64), the theorem 4.2 is proved.
\hfill\break

${\boldsymbol{Theorem \ 4.3: }}$\
Suppose that the relative anisotropic relative GLHUA electric permittivity and magnetic permeability
$\varepsilon _{p,r } (r_p )$,
$\mu _{p,r } (r_p )$,
$\varepsilon _{p,\theta } (r_p )$,
$\varepsilon _{p,\phi } (r_p )$,
$\mu _{p,\theta } (r_p )$, 
$\mu _{p,\phi } (r_p )$,   
is denoted by (53), then 

\begin{equation}
\begin{array}{l}
 \mathop {\lim }\limits_{r_p  \to R_1 } E_{p,\theta } (\vec r_p ) = 0, \\ 
 \mathop {\lim }\limits_{r_p  \to R_1 } E_{p,\phi } (\vec r_p ) = 0, \\ 
 \end{array}
\end{equation}

\begin{equation}
\begin{array}{l}
 \mathop {\lim }\limits_{r_p  \to R_1 } H_{p,\theta } (\vec r_p ) = 0, \\ 
 \mathop {\lim }\limits_{r_p  \to R_1 } H_{p,\phi } (\vec r_p ) = 0, \\ 
 \end{array}
\end{equation}

${\boldsymbol{Proof: }}$\

Similar with proof process of theorem 4.1, by GL transform (40)-(42), the angular electromagnetic wave field in physical GLHUA outer annular layer Cloak, $R_1  \le r_p  \le R_2 $, is induced from the GLHUA angular electromagnetic wave in virtual invisible sphere.
 \begin{equation}
\begin{array}{l}
 E_{p,\theta } (\vec r_p ) = E_{p,\theta } (\vec r_p (r)) \\ 
  = \frac{1}{{r_p }}rE_\theta  (\vec r), \\ 
 \end{array}
\end{equation}
Based on the theorem 6.3 in [1] and by direct calculation, we have

\begin{equation}
\begin{array}{l}
 \mathop {\lim }\limits_{r_p  \to R_1 } E_{p,\theta } (\vec r_p ) \\ 
  = \mathop {\lim }\limits_{r \to 0} E_{p,\theta } (\vec r_p (r)) \\ 
  = \mathop {\lim }\limits_{r \to 0} \frac{1}{{r_p }}rE_\theta  (\vec r) \\ 
  = \frac{1}{{R_1 }}\mathop {\lim }\limits_{r \to 0} rE_\theta  (\vec r) = 0, \\ 
 \end{array}
\end{equation}

The first limitation of (67) is proved, similarly, we can prove the second limitation of (67) and
limitation equations in (68), the theorem 4.3 is proved.
\hfill\break

${\boldsymbol{Theorem \ 4.4: }}$\
Suppose that the relative anisotropic relative GLHUA electric permittivity and magnetic permeability
$\varepsilon _{p,r } (r_p )$,
$\mu _{p,r } (r_p )$,
$\varepsilon _{p,\theta } (r_p )$,
$\varepsilon _{p,\phi } (r_p )$,
$\mu _{p,\theta } (r_p )$, 
$\mu _{p,\phi } (r_p )$,   
is denoted by (53), then 

\begin{equation}
\begin{array}{l}
 \mathop {\lim }\limits_{r_p  \to R_1 } \frac{1}{{\varepsilon _{p,\theta } }}\frac{\partial }{{\partial r_p }}E_{p,\theta } (\vec r_p ) = 0, \\ 
 \mathop {\lim }\limits_{r_p  \to R_1 } \frac{1}{{\varepsilon _{p,\theta } }}\frac{\partial }{{\partial r_p }}E_{p,\theta } (\vec r_p ) = 0, \\ 
 \end{array}
\end{equation}

\begin{equation}
\begin{array}{l}
 \mathop {\lim }\limits_{r_p  \to R_1 } \frac{1}{{\mu _{p,\theta } }}\frac{\partial }{{\partial r_p }}H_{p,\theta } (\vec r_p ) = 0, \\ 
 \mathop {\lim }\limits_{r_p  \to R_1 } \frac{1}{{\mu _{p,\theta } }}\frac{\partial }{{\partial r_p }}H_{p,\theta } (\vec r_p ) = 0, \\ 
 \end{array}
\end{equation}
\hfill\break

${\boldsymbol{Proof: }}$\ 

Similar with the proof process of theorem 4.2, by direct calculation, the derivative of the angular electric wave in GLHUA physical outer annular layer cloak,$R_1  \le r_p  \le R_2 $, is induced by GL transform (40)-(42),

\begin{equation}
\begin{array}{l}
 \frac{1}{{\varepsilon _{p,\theta } }}\frac{\partial }{{\partial r_p }}E_{p,\theta } (\vec r_p ) =  \\ 
  = \frac{1}{{\varepsilon _{p,\theta } }}\frac{{\partial r}}{{\partial r_p }}\frac{\partial }{{\partial r}}E_{p,\theta } (\vec r_p (r)) \\ 
  = \frac{1}{{\varepsilon _\theta  }}\frac{1}{{r_p }}\frac{\partial }{{\partial r}}rE_\theta  (\vec r) -  \\ 
  - \frac{1}{{\varepsilon _{p,\theta } }}\frac{{\partial r}}{{\partial r_p }}\frac{1}{{r_p ^2 }}\frac{{\partial r_p }}{{\partial r}}rE_\theta  (\vec r) \\ 
  = \frac{1}{{r_p }}\frac{1}{{\varepsilon _\theta  }}\frac{\partial }{{\partial r}}rE_\theta  (\vec r) -  \\ 
  - \frac{1}{{\varepsilon _{p,\theta } }}\frac{1}{{r_p ^2 }}rE_\theta  (\vec r), \\ 
 \end{array}
\end{equation}
Based on the theorem 6.3 and theorem 6.4 in [1], we have
\begin{equation}
\begin{array}{l}
 \mathop {\lim }\limits_{r_p  \to R_1 } \frac{1}{{\varepsilon _{p,\theta } }}\frac{\partial }{{\partial r_p }}E_{p,\theta } (\vec r_p ) =  \\ 
  = \mathop {\lim }\limits_{r \to 0} \frac{1}{{\varepsilon _{p,\theta } }}\frac{\partial }{{\partial r_p }}E_{p,\theta } (\vec r_p (r)) \\ 
  = \mathop {\lim }\limits_{r \to 0} \frac{1}{{r_p }}\frac{1}{{\varepsilon _\theta  }}\frac{\partial }{{\partial r}}rE_\theta  (\vec r) -  \\ 
  - \mathop {\lim }\limits_{r \to 0} \frac{1}{{\varepsilon _{p,\theta } }}\frac{1}{{r_p ^2 }}rE_\theta  (\vec r) \\ 
  = \frac{1}{{R_1 }}\mathop {\lim }\limits_{r \to 0} \frac{1}{{\varepsilon _\theta  }}\frac{\partial }{{\partial r}}rE_\theta  (\vec r) -  \\ 
  - \frac{1}{{R_1 ^2 }}\mathop {\lim }\limits_{r \to 0} \frac{1}{{\varepsilon _{p,\theta } }}rE_\theta  (\vec r) = 0, \\ 
 \end{array}
\end{equation}

The first limitation equation of (71) is proved, similarly, we can prove the second limitation equation of (71) and two limitation equations in (72), the theorem 4.4 is proved.

\section{The incident electromagnetic wave excited outside GLHUA cloak can not be disturbed by the cloak and can not propagation penetrate into the sphere $r < R_1$}

In the section 2, we proposed relative parameters of GLHUA  outer annular layer cloak $R_1  < r \le R_2 $, and GKHUA inner layer cloak, $R_0  < r \le R_1 $. In this section, we will prove that the incident electromagnetic wave excited in outside of GLHUA double layer cloak can not propagation penetrate into the sphere 
$r  < R_1$, and the outside incident wave can not be disturbed by the cloak. The incident electromagnetic wave excited in outside of GLHUA double layer cloak is propagation without infinite speed and without exceeding light speed.

In the section 6 in paper [1] , we proposed GLHUA sphere  pre cloak material condition (6.1)-(6.4) in invisible virtual sphere $ r  \le R_2 $. We did detailed theoretical investigating analysis for electromagnetic wave field propagation and proved theorem 6.1-6.4 in the GLHUA invisible virtual sphere $ r  \le R_2 $, use notation $\vec r = (r,\theta ,\phi )$ to denote coordinate. In the section 3, we find GL transform which mapping the physical outer annular layer $R_1  < r \le R_2 $,  to GLHUA virtual sphere $r \le R_2$, we create the GLHUA electromagnetic relative material parameters in the physical outer sphere annular layer $R_1  < r \le R_2 $. In section 4, we proved theorem 4.1-4.4 for electromagnetic wave propagation and going to zero at inner boundary $r = R_1$ in the physical GLHUA outer layer invisible cloak with parameter not less 1 and without exceeding light speed propagation. The notation $\vec r_p  = (r_p ,\theta ,\phi )$ is used to denote coordinate in physical space in section 3 and section 4. In this section, we rigorously prove the theorem for GLHUA invisible cloak in physical outer annular layer $R_1  < r \le R_2 $ and inner layer $R_0  < r \le R_1 $ with concealment $r < R_0$, the notation $\vec r  = (r ,\theta ,\phi )$ is also re used to denote coordinate and vector in physical outer annular layer cloak and inner layer cloak and concealment. Note that coordinate $\vec r_p  = (r_p ,\theta ,\phi )$  in theorem 4.1 - theorem 4.4 should be coordinate $\vec r  = (r ,\theta ,\phi )$ in this section.
\hfill\break

${\boldsymbol{Theorem \ 5.1: }}$\ 
Suppose that in the outer sphere annular layer of GLHUA double layer cloak, $R_1  < r \le R_2 $, the relative anisotropic electric permittivity and magnetic permeability parameter is proposed in (1); in the inner sphere annular layer of GLHUA double cloak,  $R_0  < r \le R_1 $, the relative anisotropic electric permittivity and magnetic permeability is proposed in  (12),  the source $r_s > R_2 $ and observer $r_o < R_1$, then 

\begin{equation}
E_r (\vec r_o ) = 0,
\end{equation}
\begin{equation}
H_r (\vec r_o ) = 0,
\end{equation}

${\boldsymbol{Proof: }}$\ 
Let radial GL electric wave $E(\vec r) = r^2 E_r (\vec r)$, $E_r (\vec r) $ is the radial electric wave. By theorem 4.1, the GL electric wave interface condition on the sphere surface,
$ r = R_1 $,

\begin{equation}
E(R_1 ^ -  ,\theta ,\phi ) = E(R_1 ^ +  ,\theta ,\phi ) = 0,
\end{equation}

By theorem 4.2, the radial GL electric wave derivative interface condition on the sphere surface, 
$r = R_1$,

\begin{equation}
\begin{array}{l}
 \frac{1}{{\varepsilon _\theta  (R_1 ^ -  )}}\frac{\partial }{{\partial r}}E(R_1 ^ -  ,\theta ,\phi ) =  \\ 
  = \frac{1}{{\varepsilon _\theta  (R_1 ^ +  )}}\frac{\partial }{{\partial r}}E_r (R_1 ^ +  ,\theta ,\phi ) = 0 \\ 
 \end{array}
\end{equation}
Because the sphere body $r \le R_1$ includes the GLHUA inner layer cloak material parameters in (12) in $R_0  < r \le R_1 $ and free space concealment $ r < R_0 $, the radial GL electric wave field does satisfy  

\begin{equation}
\begin{array}{l}
 \frac{\partial }{{\partial r}}\frac{1}{{\varepsilon _\theta  }}\frac{\partial }{{\partial r}}E +  \\ 
  + \frac{1}{{r^2 }}\frac{1}{{\sin \theta }}\frac{\partial }{{\partial \theta }}\sin \theta \frac{{\partial E}}{{\partial \theta }} +  \\ 
  + \frac{1}{{r^2 }}\frac{1}{{\sin ^2 \theta }}\frac{{\partial ^2 E}}{{\partial \phi ^2 }} + k^2 \mu _\theta  E = 0, \\ 
 \end{array}
\end{equation}

The corresponding Green��s function does satisfy the Green��s equation,

\begin{equation}
\begin{array}{l}
 \frac{\partial }{{\partial r}}\frac{1}{{\varepsilon _\theta  }}\frac{\partial }{{\partial r}}G(\vec r,\vec r_o ) \\ 
  + \frac{1}{{r^2 }}\frac{1}{{\sin \theta }}\frac{\partial }{{\partial \theta }}\sin \theta \frac{\partial }{{\partial \theta }}G(\vec r,\vec r_o ) +  \\ 
  + \frac{1}{{r^2 }}\frac{1}{{\sin ^2 \theta }}\frac{{\partial ^2 }}{{\partial \phi ^2 }}G(\vec r,\vec r_o ) \\ 
  + k^2 \mu _\theta  G(\vec r,\vec r_0 ) = \delta (\vec r - \vec r_0 ), \\ 
 \end{array}
\end{equation}

where $\varepsilon _\theta   = \mu _\theta  $ is shown in (12) in the GLHUA inner cloak $R_0  < r \le R_1 $, $\varepsilon _\theta   = \mu _\theta   = 1$ in 
concealment free space $ r < R_0$. Because $r_o < R_1$, similar with proof of theorem 4.1 and 4.2
and theorem 6.1 and 6.2 in paper [1], in the inner layer cloak $R_0  < r \le R_1 $ with GLHUA cloak relative material parameters in (12), we have 
\begin{equation}
\begin{array}{l}
 G(\vec r,\vec r_0 )|_{r = R_{_1 }^ -  }  = 0, \\ 
 \frac{1}{{\varepsilon _\theta  }}\frac{\partial }{{\partial r}}G(\vec r,\vec r_0 )|_{r = R_{_1 }^ -  }  = 0, \\ 
 \end{array}
\end{equation}
Use $G(\vec r,\vec r_0 )$  to mutiply (79), take integral on sphere $ r \le R_1$ , and using integral by part, we have

\begin{equation}
\begin{array}{l}
 \int_0^\pi  {\int_0^{2\pi } {\frac{1}{{\varepsilon _\theta  (R_1 ^ -  )}}\frac{\partial }{{\partial r}}E(R_1 ^ -  ,\theta ,\phi )G(R_1 ^ -  ,\theta ,\phi )\sin \theta d\theta d\phi } }  \\ 
  - \int_0^\pi  {\int_0^{2\pi } {\frac{\partial }{{\partial r}}E(0,\theta ,\phi )G(0,\theta ,\phi )\sin \theta d\theta d\phi } }  -  \\ 
  - \int_{S(r < R_1 )} {\frac{1}{{\varepsilon _\theta  }}\frac{\partial }{{\partial r}}E\frac{\partial }{{\partial r}}GdV}  \\ 
  + \int_{S(r < R_1 )} {\left( {\frac{1}{{r^2 }}\frac{1}{{\sin \theta }}\frac{\partial }{{\partial \theta }}\sin \theta \frac{{\partial E}}{{\partial \theta }}} \right.}  +  \\ 
 \left. { + \frac{1}{{r^2 }}\frac{1}{{\sin ^2 \theta }}\frac{{\partial ^2 E}}{{\partial \phi ^2 }}} \right)GdV \\ 
  + \int_{S(r < R_1 )} {k^2 \mu _\theta  EGdv}  = 0, \\ 
 \end{array}
\end{equation}

Use $\vec E(r)$  to multiply (80), take integral on sphere $ r\le R_1$ , and using integral by part, we have

\begin{equation}
\begin{array}{l}
 \int_0^\pi  {\int_0^{2\pi } {\frac{1}{{\varepsilon _\theta  (R_1 ^ -  )}}E(R_1 ^ -  ,\theta ,\phi )} }  
 \frac{\partial }{{\partial r}}G(R_1 ^ -  ,\theta ,\phi ,\vec r_o )\sin \theta d\theta d\phi  \\ 
  - \int_0^\pi  {\int_0^{2\pi } {E(0,\theta ,\phi )\frac{\partial }{{\partial r}}G(0,\theta ,\phi ,\vec r_o )\sin \theta d\theta d\phi } }  \\ 
  - \int_{S(r < R_1 )} {\frac{1}{{\varepsilon _\theta  (R_1 ^ -  )}}\frac{\partial }{{\partial r}}E\frac{\partial }{{\partial r}}GdV}  \\ 
  + \int_{S(r < R_1 )} {\left( {\frac{1}{{r^2 }}\frac{1}{{\sin \theta }}\frac{\partial }{{\partial \theta }}\sin \theta \frac{{\partial G}}{{\partial \theta }}} \right. + }  \\ 
  + \left. {\frac{1}{{r^2 }}\frac{1}{{\sin ^2 \theta }}\frac{{\partial ^2 G}}{{\partial \phi ^2 }}} \right)EdV \\ 
  + \int_{S(r < R_1 )} {k^2 \mu _\theta  EGdv}  = E(\vec r_o ), \\ 
 \end{array}
\end{equation}

To subtract (82) from (83), we have
\begin{equation}
\begin{array}{l}
 E(\vec r_o ) = \int_0^\pi  {\int_0^{2\pi } {\frac{1}{{\varepsilon _\theta  (R_1 ^ -  )}}\frac{\partial }{{\partial r}}E(R_1 ^ -  ,\theta ,\phi )} }  \\ 
 G(R_1 ^ -  ,\theta ,\phi )\sin \theta d\theta d\phi  \\ 
  - \int_0^\pi  {\int_0^{2\pi } {E(R_1 ^ -  ,\theta ,\phi )} }  \\ 
 \frac{1}{{\varepsilon _\theta  (R_1 ^ -  )}}\frac{\partial }{{\partial r}}G(R_1 ^ -  ,\theta ,\phi ,\vec r_o )\sin \theta d\theta d\phi , \\ 
 \end{array}
\end{equation}
Based on the interface condition (77) and (78), we have
\begin{equation}
 E(\vec r_o ) = 0,
\end{equation}
for $r_o \ne 0$,we have
\begin{equation}
E_r (\vec r_o ) = \frac{1}{{r_o ^2 }}E(\vec r_o ) = 0,
\end{equation}
for $r_o = 0$,by continuty
\begin{equation}
 E_r(\vec r_o ) = 0,
\end{equation}
The equation (75) is proved, similarly, we proved equation (75). Therefore, theorem 5.1 is proved.
\hfill\break

${\boldsymbol{Theorem \ 5.2: }}$\ 
Suppose that in the outer sphere annular layer of GLHUA double layer cloak, 
$R_1  < r \le R_2 $, the anisotropic relative electric permittivity and magnetic permeability parameter is in (1); in the inner sphere annular layer GLHUA double cloak, $R_0  < r \le R_1 $,  the relative anisotropic electric permittivity and magnetic permeability is in  (12)�� the source $r_s > R_2$  and observer $r_o < R_1$��then 

\begin{equation}
\begin{array}{l}
 E_\theta  (\vec r_o ) = 0, \\ 
 E_\phi  (\vec r_o ) = 0, \\ 
 \end{array}
 \end{equation}
\begin{equation}
\begin{array}{l}
H_\theta  (\vec r_o ) = 0, \\ 
 H_\phi  (\vec r_o ) = 0, \\ 
 \end{array}
 \end{equation}

${\boldsymbol{Proof: }}$\ 
 By theorem 4.3, the angular electric wave interface condition on the sphere surface, $r = R_1$,
 \begin{equation}
 E_\theta  (R_1 ^ -  ,\theta ,\phi ) = E_\theta  (R_1 ^ +  ,\theta ,\phi ) = 0,
 \end{equation}
By theorem 4.4, the angular electric wave derivative interface condition on the sphere surface,$r  = R_1$ ,
 \begin{equation}
\frac{\partial }{{\partial r}}E_\theta  (R_1 ^ -  ,\theta ,\phi ) = \frac{1}{{\varepsilon _\theta  }}\frac{\partial }{{\partial r}}E_\theta  (R_1 ^ +  ,\theta ,\phi ) = 0,
 \end{equation}
Because the sphere $r \le R_1$  includes GLHUA inner layer cloak material parameters in (12) in $R_0  < r \le R_1 $  and free space concealment $r < R_0$ , and based on theorem 5.1, $E_r (\vec r) = 0$ , $H_r (\vec r) = 0$  in the sphere $r < R_1$ ,
Because the source is outside sphere $r_s > R_2$, in the Sphere $0 < r \le R_1 $ , from (4) in paper [1], we have

\begin{equation}   
\begin{array}{l}
 \nabla  \cdot \vec D = \frac{1}{{r^2 }}\frac{\partial }{{\partial r}}\left( {r^2 \varepsilon _r E_r } \right) +  \\ 
  + \frac{1}{{\sin \theta r}}\frac{\partial }{{\partial \theta }}\sin \theta \varepsilon _\theta  E_\theta   + \frac{1}{{r\sin \theta }}\frac{\partial }{{\partial \phi }}\varepsilon _\phi  E_\phi   = 0 \\ 
 \end{array}
\end{equation}

The $\vec D$  is displacement electric in spherical coordinate, by definition of GL electromagnetic wave in
(18)in [1], equation (92) becomes
\begin{equation}
\frac{1}{{\sin \theta }}\frac{\partial }{{\partial \theta }}\sin \theta rE_\theta   + \frac{1}{{\sin \theta }}\frac{\partial }{{\partial \phi }}rE_\phi   =  - \frac{1}{{\varepsilon _\theta  }}\frac{{\partial E}}{{\partial r}}
\end{equation}
By Maxwell equation (1) and (18) in paper [1], we have
 \begin{equation}                      
 - \frac{1}{{\sin \theta }}\frac{\partial }{{\partial \phi }}rE_\theta   + \frac{1}{{\sin \theta }}\frac{\partial }{{\partial \theta }}\sin \theta rE_\phi   =  - i\omega \mu _0 H,
\end{equation}
Rewrite (93) and (94) as matrix equation
\begin{equation}
\left[ {\begin{array}{*{20}c}
   {\frac{1}{{\sin \theta }}\frac{\partial }{{\partial \theta }}\sin \theta } & {\frac{1}{{\sin \theta }}\frac{\partial }{{\partial \phi }}}  \\
   { - \frac{1}{{\sin \theta }}\frac{\partial }{{\partial \phi }}} & {\frac{1}{{\sin \theta }}\frac{\partial }{{\partial \theta }}\sin \theta }  \\
\end{array}} \right]\left[ {\begin{array}{*{20}c}
   {rE_\theta  }  \\
   {rE_\phi  }  \\
\end{array}} \right] = \left[ {\begin{array}{*{20}c}
   { - \frac{1}{{\varepsilon _\theta  }}\frac{{\partial E}}{{\partial r}}}  \\
   { - i\omega \mu _0 H}  \\
\end{array}} \right]
\end{equation}
The adjoint GLHUA Green��s equation of equation (95) on  $[0,\pi ;0,2\pi ]$ is an angular Green equation
\begin{equation}
\begin{array}{l}
 \left[ {\begin{array}{*{20}c}
   {\frac{1}{{\sin \theta }}\frac{\partial }{{\partial \theta }}\sin \theta } & { - \frac{1}{{\sin \theta }}\frac{\partial }{{\partial \phi }}}  \\
   {\frac{1}{{\sin \theta }}\frac{\partial }{{\partial \phi }}} & {\frac{1}{{\sin \theta }}\frac{\partial }{{\partial \theta }}\sin \theta }  \\
\end{array}} \right] \\ 
 \left[ {\begin{array}{*{20}c}
   {G_{11} (\theta ,\theta ',\phi ,\phi ')} & {G_{21} (\theta ,\theta ',\phi ,\phi ')}  \\
   {G_{12} (\theta ,\theta ',\phi ,\phi ')} & {G_{22} (\theta ,\theta ',\phi ,\phi ')}  \\
\end{array}} \right] \\ 
  = \frac{1}{{\sin \theta }}\left[ {\begin{array}{*{20}c}
   {\delta (\theta ,\theta ',\phi ,\phi ')} & {}  \\
   {} & {\delta (\theta ,\theta ',\phi ,\phi ')}  \\
\end{array}} \right], \\ 
 \end{array}
\end{equation}
Where the GLHUA angular Green matrix is presented in (115)in paper [1]. Similar with proof of the theorem 6.3 in paper [1]. Based on theorem 5.1,$E_r (\vec r) = 0$ , $H_r (\vec r) = 0$ , also $\frac{1}{{\varepsilon _\theta  }}\frac{\partial }{{\partial r}}E = 0$, $\frac{1}{{\varepsilon _\theta  }}\frac{\partial }{{\partial r}}H = 0$ in the sphere $r < R_1$, therefore, we have
\begin{equation}
\begin{array}{l}
 \left[ {\begin{array}{*{20}c}
   {rE_\theta  (r,\theta ',\phi ')}  \\
   {rE_\phi  (r,\theta ',\phi ')}  \\
\end{array}} \right] =  \\ 
  = \int_0^\pi  {} \int_0^{2\pi } {} \left[ {\begin{array}{*{20}c}
   {G_{11} } & {G_{12} }  \\
   {G_{21} } & {G_{22} }  \\
\end{array}} \right]\left[ {\begin{array}{*{20}c}
   { - \frac{1}{{\varepsilon _\theta  }}\frac{{\partial E}}{{\partial r}}}  \\
   { - i\omega \mu _0 H}  \\
\end{array}} \right]\sin \theta d\theta d\phi  \\ 
  = \int_0^\pi  {} \int_0^{2\pi } {} \left[ {\begin{array}{*{20}c}
   {G_{11} } & {G_{12} }  \\
   {G_{21} } & {G_{22} }  \\
\end{array}} \right]\left[ {\begin{array}{*{20}c}
   0  \\
   0  \\
\end{array}} \right]\sin \theta d\theta d\phi  \\ 
  = \left[ {\begin{array}{*{20}c}
   0  \\
   0  \\
\end{array}} \right], \\ 
 \end{array}
\end{equation}
The (88) is proved, similarly, we can prove $H_\theta  (\vec r_o ) = 0,$,
$H_\phi  (\vec r_o ) = 0,$,  in equation (89), theorem 5.2 is proved.
\hfill\break

${\boldsymbol{Theorem \ 5.3: }}$\ 
Suppose that in the outer sphere annular layer of GLHUA double layer cloak, 
$R_1  < r \le R_2 $, the anisotropic relative electric permittivity and magnetic permeability parameter is in (1); in the inner sphere annular layer GLHUA double cloak,$R_0  < r \le R_1 $, the relative anisotropic electric permittivity and magnetic permeability is in  (12), the source $r_s  > R_2 $  and observer $r_0 < R_1$ ��then 
\begin{equation}
\vec E(\vec r_o ) = 0,
 \end{equation}
\begin{equation}
\vec H(\vec r_o ) = 0,
 \end{equation}
${\boldsymbol{Proof: }}$\ 
Summary of theorem 5.1 and theorem 5.2, we can prove (98) and (99). Theorem 5.3 is proved.
\hfill\break

${\boldsymbol{Theorem \ 5.4: }}$\ 
Suppose that in the outer sphere annular layer of GLHUA double layer cloak, $R_1  < r \le R_2 $, the anisotropic relative electric permittivity and magnetic permeability parameter is proposed in (1); in the inner sphere annular layer GLHUA double cloak, $R_0  < r \le R_1 $, the relative anisotropic electric permittivity and magnetic permeability is proposed in (12), the source $r_s > R_2$  and observer $r_o > R_2  $, then the radial electromagnetic wave equal to the incident wave

\begin{equation}
\begin{array}{l}
 E_r (\vec r_o ) = E_r^b (\vec r_o ), \\ 
 H_r (\vec r_o ) = H_r^b (\vec r_o ). \\ 
 \end{array}
 \end{equation}

${\boldsymbol{Proof: }}$\ 
 Based on theorem 5.1-5.3, if the source $r_s > R_2$ , in the sphere $r < R_1$   of GLHUA double cloak, $\vec E(\vec r) = 0$  and $\vec H(\vec r) = 0,$ because the source  $r_s > R_2$  and observer $r_o > R_2$ . In the GLHUA outer sphere annular layer cloak  $R_1  \le r \le R_2 $

 \begin{equation}
 \begin{array}{l}
 \frac{\partial }{{\partial r}}\frac{1}{{\varepsilon _\theta  }}\frac{\partial }{{\partial r}}E +  \\ 
  + \frac{1}{{r^2 }}\frac{1}{{\sin \theta }}\frac{\partial }{{\partial \theta }}\sin \theta \frac{{\partial E}}{{\partial \theta }} +  \\ 
  + \frac{1}{{r^2 }}\frac{1}{{\sin ^2 \theta }}\frac{{\partial ^2 E}}{{\partial \phi ^2 }} + k^2 \mu _\theta  E = 0, \\ 
 \end{array}
 \end{equation}

the corresponding Green��s function satisfy the following Green��s equation with virtual source term $r_o > R_2$,

 \begin{equation}
\begin{array}{l}
 \frac{\partial }{{\partial r}}\frac{\partial }{{\partial r}}G(\vec r,\vec r') +  \\ 
  + \frac{1}{{r^2 }}\frac{1}{{\sin \theta }}\frac{\partial }{{\partial \theta }}\sin \theta \frac{\partial }{{\partial \theta }}G(\vec r,\vec r_o ) \\ 
  + \frac{1}{{r^2 }}\frac{1}{{\sin ^2 \theta }}\frac{{\partial ^2 }}{{\partial \phi ^2 }}G(\vec r,\vec r_o ) \\ 
  + k^2 G(\vec r,\vec r_o ) = \delta (\vec r - \vec r_o ), \\ 
 \end{array}
 \end{equation}
Similar with proof process of integral equation (51), from (101) and (102), we have integral equation

\begin{equation}
\begin{array}{l}
 E(\vec r_o ) = E^b (\vec r_o ) \\ 
  - \int\limits_{S(R_1  \le r \le R_2 )} {\left( {1 - \frac{1}{{\varepsilon _\theta  }}} \right)\frac{\partial }{{\partial r}}G(\vec r,\vec r_o )\frac{\partial }{{\partial r}}E(r)dV}  \\ 
  + \int\limits_{S(R_1  \le r \le R_2 )} {k^2 \left( {1 - \mu _\theta  } \right)G(\vec r,\vec r_o )EdV} . \\ 
 \end{array}
\end{equation}

Let the scattering integral be
\begin{equation}
\begin{array}{l}
 ScatI =  \\ 
  - \int\limits_{S(R_1  \le r \le R_2 )} {\left( {1 - \frac{1}{{\varepsilon _\theta  }}} \right)\frac{\partial }{{\partial r}}G(\vec r,\vec r_o )\frac{\partial }{{\partial r}}E(r)dV}  \\ 
  + \int\limits_{S(R_1  \le r \le R_2 )} {k^2 \left( {1 - \mu _\theta  } \right)G(\vec r,\vec r_o )EdV} . \\ 
 \end{array}
\end{equation}

By calculation and integral by parts, the scattering term becomes
\begin{equation}
\begin{array}{l}
 ScatI =  \\ 
  - \int\limits_{S(R_1  \le r \le R_2 )} {\frac{\partial }{{\partial r}}\left( {\frac{\partial }{{\partial r}}G(\vec r,\vec r_o )E(r)} \right)dV}  +  \\ 
  + \int\limits_{S(R_1  \le r \le R_2 )} {\left( {\frac{\partial }{{\partial r}}\frac{\partial }{{\partial r}}G(\vec r,\vec r_o )} \right)E(r)dV}  \\ 
  + \int\limits_{S(R_1  \le r \le R_2 )} {\left( {\frac{1}{{r^2 }}\frac{1}{{\sin \theta }}\frac{\partial }{{\partial \theta }}\sin \theta \frac{{\partial G(\vec r,\vec r_o )}}{{\partial \theta }}} \right. + }  \\ 
 \left. { + \frac{1}{{r^2 }}\frac{1}{{\sin ^2 \theta }}\frac{{\partial ^2 G(\vec r,\vec r_o )}}{{\partial \phi ^2 }}} \right)EdV \\ 
  + \int\limits_{S(R_1  \le r \le R_2 )} {k^2 G(\vec r,\vec r_o )EdV}  \\ 
  + \int\limits_{S(R_1  \le r \le R_2 )} {\frac{\partial }{{\partial r}}\left( {G(\vec r,\vec r_o )\frac{1}{{\varepsilon _\theta  }}\frac{\partial }{{\partial r}}E(r)} \right)dV}  \\ 
  - \int\limits_{S(R_1  \le r \le R_2 )} {\left( {\frac{\partial }{{\partial r}}\frac{1}{{\varepsilon _\theta  }}\frac{\partial }{{\partial r}}E(r)} \right)G(\vec r,\vec r_o )dV}  \\ 
  - \int\limits_{S(R_1  \le r \le R_2 )} {\left( {\frac{1}{{r^2 }}\frac{1}{{\sin \theta }}\frac{\partial }{{\partial \theta }}\sin \theta \frac{{\partial E}}{{\partial \theta }}} \right. + }  \\ 
 \left. { + \frac{1}{{r^2 }}\frac{1}{{\sin ^2 \theta }}\frac{{\partial ^2 E}}{{\partial \phi ^2 }}G(\vec r,\vec r_o )} \right)dV \\ 
  - \int\limits_{S(R_1  \le r \le R_2 )} {k^2 \mu _\theta  G(\vec r,\vec r_o )EdV} . \\ 
 \end{array}
\end{equation}

The scattering term is reduced to
\begin{equation}
\begin{array}{l}
 ScatI =  - \int\limits_{S(R_1  \le r \le R_2 )} {} \frac{\partial }{{\partial r}}(\frac{\partial }{{\partial r}}G(\vec r,\vec r_o )E(r))dV \\ 
  + \int\limits_{S(R_1  \le r \le R_2 )} {} \frac{\partial }{{\partial r}}\left( {} \right.G(\vec r,\vec r_o )\frac{1}{{\varepsilon _\theta  }}\frac{\partial }{{\partial r}}E(r)\left. {} \right)dV, \\ 
 \end{array}
\end{equation}

\begin{equation}
\begin{array}{l}
 ScatI =  - \int_0^\pi  {\int_0^{2\pi } {\left. {\frac{\partial }{{\partial r}}G(\vec r,\vec r_o )E(r)} \right|_{r = R_2 } \sin \theta d\theta d\phi } }  \\ 
  + \int_0^\pi  {\int_0^{2\pi } {\left. {\frac{\partial }{{\partial r}}G(\vec r,\vec r_o )E(r)} \right|_{r = R_1 } \sin \theta d\theta d\phi } }  \\ 
  + \int_0^\pi  {\int_0^{2\pi } {G(\vec r,\vec r_o )\left. {\frac{\partial }{{\partial r}}E(r)} \right|_{r = R_2 } \sin \theta d\theta d\phi } }  \\ 
  - \int_0^\pi  {\int_0^{2\pi } {G(\vec r,\vec r_o )\frac{1}{{\varepsilon _\theta  }}\left. {\frac{\partial }{{\partial r}}E(r)} \right|_{r = R_1 } \sin \theta d\theta d\phi } }  \\ 
 \end{array}
 \end{equation}    
 
\begin{equation} 
\begin{array}{l}
 ScatI =  \\ 
  =  - \int_0^\pi  {\int_0^{2\pi } {\left. {\frac{\partial }{{\partial r}}G(\vec r,\vec r_o )E(r)} \right|_{r = R_2 } \sin \theta d\theta d\phi } }  \\ 
  + \int_0^\pi  {\int_0^{2\pi } {G(\vec r,\vec r_o )\left. {\frac{\partial }{{\partial r}}E(r)} \right|_{r = R_2 } \sin \theta d\theta d\phi } }  = 0 \\ 
 \end{array}
 \end{equation} 

we have
 \begin{equation} 
E(\vec r_o ) = E^b (\vec r_o ),                                                             
 \end{equation} 

\begin{equation} 
\begin{array}{l}
 E_r (\vec r_o )r^2  = E(\vec r_o ) \\ 
  = E^b (\vec r_o ) = E_r^b (\vec r_o )r^2 , \\ 
 \end{array}
 \end{equation} 
 \begin{equation} 
E_r(\vec r_o ) = {E^b}_r (\vec r_o ),                                                             
 \end{equation}
First part of (100) is proved.  Similarly, we can prove
\begin{equation} 
H(\vec r_o ) = H^b (\vec r_o ),                                                             
 \end{equation}
and
\begin{equation} 
H_r(\vec r_o ) = {H^b}_r (\vec r_o ),                                                             
 \end{equation} 
 in (100). Theorem 5.4 is proved.
\hfill\break

${\boldsymbol{Theorem \ 5.5: }}$ \ 
  Suppose that in the outer sphere annular layer of GLHUA double layer cloak, $R_1  < r \le R_2 $, the anisotropic relative electric permittivity and magnetic permeability parameter is proposed in (1); in the inner sphere annular layer GLHUA double cloak, $R_0  < r \le R_1 $, the relative anisotropic electric permittivity and magnetic permeability is proposed in (12), the source $ r_s > R_2$ and observer $ r_o > R_2$, then

\begin{equation} 
 \begin{array}{l}
 E_\phi  (\vec r) = E_\phi ^b (\vec r), \\ 
 E_\phi  (\vec r) = E_\phi ^b (\vec r), \\ 
 \end{array}
\end{equation} 
\begin{equation} 
 \begin{array}{l}
 H_\phi  (\vec r) = H_\phi ^b (\vec r), \\ 
 H_\phi  (\vec r) = H_\phi ^b (\vec r), \\ 
 \end{array}
\end{equation}

${\boldsymbol{Proof: }}$\ 
 Because the source is outside sphere $r_s > R_2$, $\vec r$ is in outside of the sphere with radius $R_2$, i.e. $r > R_2$, from (4)in paper [1], we have

\begin{equation}
\begin{array}{l}
 \nabla  \cdot \vec D = \frac{1}{{r^2 }}\frac{\partial }{{\partial r}}\left( {r^2 E_r } \right) +  \\ 
  + \frac{1}{{\sin \theta r}}\frac{\partial }{{\partial \theta }}\sin \theta E_\theta   + \frac{1}{{r\sin \theta }}\frac{\partial }{{\partial \phi }}E_\phi   = q \\ 
 \end{array}
\end{equation}

In free space without cloak

 \begin{equation}
\begin{array}{l}
 \nabla  \cdot \vec D^b  = \frac{1}{{r^2 }}\frac{\partial }{{\partial r}}\left( {r^2 E_r ^b } \right) +  \\ 
  + \frac{1}{{\sin \theta r}}\frac{\partial }{{\partial \theta }}\sin \theta E_\theta  ^b  + \frac{1}{{r\sin \theta }}\frac{\partial }{{\partial \phi }}E_\phi  ^b  = q \\ 
 \end{array}
\end{equation}

Subtract (116) from (117), we have

\begin{equation}
\begin{array}{l}
 \nabla  \cdot (\vec D - \vec D^b ) = \frac{1}{{r^2 }}\frac{\partial }{{\partial r}}\left( {r^2 (E_r  - E_r ^b )} \right) \\ 
  + \frac{1}{{\sin \theta r}}\frac{\partial }{{\partial \theta }}\sin \theta (E_\theta   - E_\theta  ^b ) +  \\ 
  + \frac{1}{{r\sin \theta }}\frac{\partial }{{\partial \phi }}(E_\phi   - E_\phi  ^b ) = 0, \\ 
 \end{array}
\end{equation}

$\vec D$  is displacement electric in spherical coordinate. Based on theorem 5.4, when the source $r_s > R_2$  and observer $r_o > R_2$, then $E(\vec r) = E^b (\vec r)$, $H(\vec r) = H^b (\vec r)$ , by definition of GL electromagnetic wave (18), equation (118) becomes

\begin{equation}
\frac{1}{{\sin \theta r}}\frac{\partial }{{\partial \theta }}\sin \theta (E_\theta   - E_\theta  ^b ) + \frac{1}{{r\sin \theta }}\frac{\partial }{{\partial \phi }}(E_\phi   - E_\phi  ^b ) = 0,
\end{equation}
By Maxwell equation (1) and (18)in [1], we have
 \begin{equation}                        
  - \frac{1}{{\sin \theta }}\frac{\partial }{{\partial \phi }}(rE_\theta   - rE_\theta  ^b ) + \frac{1}{{\sin \theta }}\frac{\partial }{{\partial \theta }}\sin \theta (rE_\phi   - rE_\phi  ^b ) = 0,
\end{equation}

Rewrite (119) and (120) as matrix equation
\begin{equation}
\left[ {\begin{array}{*{20}c}
   {\frac{1}{{\sin \theta }}\frac{\partial }{{\partial \theta }}\sin \theta } & {\frac{1}{{\sin \theta }}\frac{\partial }{{\partial \phi }}}  \\
   { - \frac{1}{{\sin \theta }}\frac{\partial }{{\partial \phi }}} & {\frac{1}{{\sin \theta }}\frac{\partial }{{\partial \theta }}\sin \theta }  \\
\end{array}} \right]\left[ {\begin{array}{*{20}c}
   {rE_\theta   - rE_\theta  ^b }  \\
   {rE_\phi   - rE_\phi  ^b }  \\
\end{array}} \right] = \left[ {\begin{array}{*{20}c}
   0  \\
   0  \\
\end{array}} \right],
\end{equation}
The adjoint GLHUA angular Green��s equation of equation (121) on $[0,\pi ;0,2\pi ]$  is
\begin{equation}
\begin{array}{l}
 \left[ {\begin{array}{*{20}c}
   {\frac{1}{{\sin \theta }}\frac{\partial }{{\partial \theta }}\sin \theta } & { - \frac{1}{{\sin \theta }}\frac{\partial }{{\partial \phi }}}  \\
   {\frac{1}{{\sin \theta }}\frac{\partial }{{\partial \phi }}} & {\frac{1}{{\sin \theta }}\frac{\partial }{{\partial \theta }}\sin \theta }  \\
\end{array}} \right] \\ 
 \left[ {\begin{array}{*{20}c}
   {G_{11} (\theta ,\theta ',\phi ,\phi ')} & {G_{21} (\theta ,\theta ',\phi ,\phi ')}  \\
   {G_{12} (\theta ,\theta ',\phi ,\phi ')} & {G_{22} (\theta ,\theta ',\phi ,\phi ')}  \\
\end{array}} \right] \\ 
  = \frac{1}{{\sin \theta }}\left[ {\begin{array}{*{20}c}
   {\delta (\theta ,\theta ',\phi ,\phi ')} & {}  \\
   {} & {\delta (\theta ,\theta ',\phi ,\phi ')}  \\
\end{array}} \right], \\ 
 \end{array}
 \end{equation}   
Where the GLHUA angular Green matrix is proposed in (115) in paper [1]. Similar with proof of the theorem 6.3 in paper [1]. Based on theorem 5.4, $E(\vec r) = E^b (\vec r),$ $H(\vec r) = H^b (\vec r),$  in the outside of the sphere,i.e.,
$r > R_2$ , we have
 \begin{equation}   
\begin{array}{l}
 \left[ {\begin{array}{*{20}c}
   {rE_\theta  (r,\theta ',\phi ') - rE_\theta  ^b (r,\theta ',\phi ')}  \\
   {rE_\phi  (r,\theta ',\phi ') - rE_\phi  ^b (r,\theta ',\phi ')}  \\
\end{array}} \right] =  \\ 
  = \int_0^\pi  {} \int_0^{2\pi } {} \left[ {\begin{array}{*{20}c}
   {G_{11} } & {G_{12} }  \\
   {G_{21} } & {G_{22} }  \\
\end{array}} \right]\left[ {\begin{array}{*{20}c}
   0  \\
   0  \\
\end{array}} \right]\sin \theta d\theta d\phi  = \left[ {\begin{array}{*{20}c}
   0  \\
   0  \\
\end{array}} \right], \\ 
 \end{array}
 \end{equation}   
We proved (114) in the first part of the theorem 5.5, similarly, we can prove (115). Therefore, the theorem 5.5 is proved.
\hfill\break

${\boldsymbol{Theorem \ 5.6: }}$\ 
Theorem 5.6. Suppose that in the outer sphere annular layer of GLHUA double layer cloak, $R_1  < r \le R_2 $, the anisotropic relative electric permittivity and magnetic permeability parameter is proposed in (1); in the inner sphere annular layer GLHUA double cloak,$R_0  < r \le R_1 $, the relative anisotropic electric permittivity and magnetic permeability is proposed in (12), the source $r_s > R_2$ and observer $r_o > R_2$, then 

\begin{equation}   
\begin{array}{l}
 \vec E(\vec r) = \vec E^b (\vec r), \\ 
 \vec H(\vec r) = \vec H^b (\vec r), \\ 
 \end{array}
\end{equation}   

${\boldsymbol{Proof: }}$\
Summary of theorem 5.4 and theorem 5.5, we obtained the proof of (124). Theorem 5.6 is proved.

\section{The incident electromagnetic wave excited inside GLHUA cloak concealment $r_s < R_0$ can not be disturbed by the cloak and can not propagate to $r > R_1$ , i.e. can not propagate to the outside of the inner annular layer cloak}

In the section 1, we proposed GLHUA double layer invisible cloak with outer physical sphere annular layer, $R_1  < r \le R_2 $, inner layer, $R_0  < r \le R_1 $   and concealment $r < R_0 $. In this section, in theorem 6.3, we will prove that the incident electromagnetic wave excited in inside of GLHUA cloak concealment $r_s < R_0$  can not propagation to  $r > R_1$ , i.e. can not propagation to outside of inner layer; in theorem 6.6 we prove that  the incident wave excited in inside concealment can not be disturbed by the cloak; The incident electromagnetic wave excited in inside concealment $r_s < R_0$  of GLHUA double layer cloak is propagation without infinite speed and without exceeding light speed.

The following theorem 6.1 - 6.6 for the inner layer cloak are corresponding to theorem 5.1 - 5.6
for the outer layer cloak. Corresponding to the GLHUA pre cloak material conditions (6.1)-(6.4) in invisible sphere $r \le R_2$ and the theorem 6.1 to 6.4 in the paper [1], we propose corresponding GLHUA pre cloak conditions in GLHUA invisible outside sphere $r > R_0$  and theorems. Corresponding to create GLHUA outer layer cloak material parameter (1) for $R_1  < r \le R_2 $    in the subsection 1.4, we create GL inner layer cloak material parameter (12) for $R_0  < r \le R_1 $   ; and prove that in the GLHUA inner layer cloak ,$R_0  < r \le R_1 $,    with material parameter (12) not less than 1 and their continuous across $r=R_0$, in particular, we prove that when r going to the outer boundary of inner layer, $r=R_1$, the electromagnetic wave and their derivative divided by $\varepsilon _\theta  $ are going to zero. Based on these theorem we can prove that the following Theorem 6.1.to 6.4.  Next, we propose these theorems and omit their longer proof. The proof idea of this theorem is that for $\alpha  = 1$, we rewrite the relative angular parameter in (12) as 
\begin{equation}  
\begin{array}{l}
 \varepsilon _\theta   = \varepsilon _\phi   = \mu _\theta   = \mu _\phi   =  \\ 
  = \frac{1}{2}\left( {\left( {\frac{{\frac{1}{r} - \frac{1}{{R_1 }}}}{{\frac{1}{{R_0 }} - \frac{1}{{R_1 }}}}} \right) + \left( {\frac{{\frac{1}{{R_0 }} - \frac{1}{{R_1 }}}}{{\frac{1}{r} - \frac{1}{{R_1 }}}}} \right)} \right), \\ 
 R_0  < r \le R_1 , \\ 
 \end{array}
\end{equation}  
Let us to do transform 
 \begin{equation}
r_q  = \frac{1}{r}
\end{equation}

the GL electric equation is translated to 

\begin{equation}
\begin{array}{l}
 \frac{{dr_q }}{{dr}}\frac{\partial }{{\partial r_q }}\frac{1}{{\left( {\mu _\theta  } \right)}}\frac{{dr_q }}{{dr}}\frac{{\partial \left( H \right)}}{{\partial r_q }} +  \\ 
  + \frac{1}{{r_q ^2 \frac{{r^2 }}{{r_q ^2 }}\sin \theta }}\frac{\partial }{{\partial \theta }}\sin \theta \frac{{\partial H_{} }}{{\partial \theta }} +  \\ 
  + \frac{1}{{r_q ^2 \frac{{r^2 }}{{r_q ^2 }}\sin ^2 \theta }}\frac{{\partial H}}{{\partial \phi ^2 }} + k^2 \varepsilon _\theta  H = 0 \\ 
 \end{array}
\end{equation}

\begin{equation}
\begin{array}{l}
 \frac{\partial }{{\partial r_q }}\frac{1}{{\mu _{q\theta } }}\frac{{\partial \left( H \right)}}{{\partial r_q }} +  \\ 
  + \frac{1}{{r_q ^2 \mu _{qr} }}\frac{\partial }{{\partial \theta }}\sin \theta \frac{{\partial H_{} }}{{\partial \theta }} +  \\ 
  + \frac{1}{{r_q ^2 \mu _{qr} \sin ^2 \theta }}\frac{{\partial H}}{{\partial \phi ^2 }} + k^2 \varepsilon _{q\theta } H = 0 \\ 
 \end{array}
\end{equation}

\begin{equation}
\frac{1}{{R_1 }} \le r_q  \le \frac{1}{{R_0 }},
\end{equation}
\begin{equation}
\begin{array}{l}
 \mu _{q\theta }  = \mu _\theta  \frac{1}{{r_q ^2 }}, \\ 
 \mu _{qr}  = \frac{{r^2 }}{{r_q ^2 }}\frac{1}{{r^2 }} = \frac{1}{{r_q ^2 }}, \\ 
\end{array}
\end{equation}

Because $r_{qs}  = \frac{1}{{r_s }}$, incident electric wave
\begin{equation}
\begin{array}{l}
 E^b (r,\theta ,\phi ,r_s ,\theta _s ,\phi _s ) \\ 
  = E^b (\frac{1}{{r_q }},\theta ,\phi ,\frac{1}{{r_{qs} }},\theta _s ,\phi _s ), \\ 
 \end{array}
\end{equation}
\begin{equation}
\mathop {\lim }\limits_{r_q  \to 0} E^b (\frac{1}{{r_q }},\theta ,\phi ,\frac{1}{{r_{qs} }},\theta _s ,\phi _s ) = 0,
\end{equation}

Similar with proof of the theorem 4.1, 4.2, 4.3, 4.4 in this paper, we can prove that the electromagnetic field wave
excited by source in $r_{qs}  > \frac{1}{{R_0 }}$ is propagation smoothly cross boundary $r_q  = \frac{1}{{R_0 }}$ and enter annular layer $\frac{1}{{R_1 }} \le r_q  \le \frac{1}{{R_0 }}$, when $r_q$ decreasing and $r_q  \to \frac{1}{{R_1 }}$, the electromagnetic field wave is going to zero. Put back to the $\vec r$ space in the GLHUA inner annular layer cloak domain, $R_0  < r \le R_1 $, with GLHUA relative parameter in (12), by the above proof, we prove that the electromagnetic field wave excited by source in $r_s < R_0$ in concealment is propagation smoothly across boundary $r=R_0$ and enter GLHUA inner annular layer ,$R_0  < r \le R_1 $ , when $r$ increasing and $r \to R_1 $, the electromagnetic field wave is going to zero. Therefore, similar with theorem 5.1 to 5.6 for electromagnetic field wave propagation in the GLHUA outer layer cloak,$R_1  < r \le R_2 $   . We can prove theorem 6.1 to 6.6 for electromagnetic field wave propagation in the GLHUA inner layer line cloak,$R_0  < r \le R_1 $   . .
 \hfill\break

${\boldsymbol{Theorem \ 6.1: }}$\ 
  Suppose that in the outer sphere annular layer of GLHUA double layer cloak, $R_1  < r \le R_2 $, the anisotropic relative electric permittivity and magnetic permeability parameter is proposed in (1); in the inner sphere annular layer of GLHUA double cloak, $R_0  < r \le R_1 $, the relative anisotropic electric permittivity and magnetic permeability is proposed in  (12), the source $r_s < R_0$ and observer $r_o > R_1$, then radial electromagnetic wave is zero
\begin{equation}
E_r (\vec r_o ) = 0,H_r (\vec r_o ) = 0,
\end{equation}
\hfill\break
${\boldsymbol{Theorem \ 6.2: }}$\ 
Suppose that in the outer sphere annular layer of GLHUA double layer cloak, $R_1  < r \le R_2 $, the anisotropic relative electric permittivity and magnetic permeability parameter is proposed in (1); in the inner sphere annular layer GLHUA double cloak, $R_0  < r \le R_1 $, the relative anisotropic electric permittivity and magnetic permeability is proposed in (12), the source $r_s < R_0$ and observer $r_o > R_1$��then angular electromagnetic wave is zero
\begin{equation}
E_\theta  (\vec r_o ) = 0,E_\phi  (\vec r_o ) = 0,
\end{equation}
\begin{equation}
H_\theta  (\vec r_o ) = 0,H_\phi  (\vec r_o ) = 0,
\end{equation}
\hfill\break

${\boldsymbol{Theorem \ 6.3: }}$\ 
 Suppose that in the outer sphere annular layer of GLHUA double layer cloak, $R_1  < r \le R_2 $, the anisotropic relative electric permittivity and magnetic permeability parameter is proposed in (1); in the inner sphere annular layer GLHUA double cloak, $R_0  < r \le R_1 $, the relative anisotropic electric permittivity and magnetic permeability is proposed in  (12), the source $r_s < R_0$ and observer $r_o > R_1$, then the electromagnetic wave is zero

\begin{equation}
{\vec E} (\vec r_o ) = 0,{\vec H} (\vec r_o ) = 0,
\end{equation}
                                                    
Summary of theorem 6.1 and theorem 6.2, we can prove (136). Theorem 6.3 is proved. The theorem shown that the incident electromagnetic wave excited in the concealment $r_s < R_0$  can not propagate to $r > R_1 $, i.e. the incident electromagnetic wave excited in the concealment $r_s < R_0$  can not propagate to outside of the inner layer.

\hfill\break
${\boldsymbol{Theorem \ 6.4: }}$\ 
Suppose that in the outer sphere annular layer of GLHUA double layer cloak, $R_1  < r \le R_2 $,the anisotropic relative electric permittivity and magnetic permeability parameter is proposed in (1); in the inner sphere annular layer GLHUA double cloak, $R_0  < r \le R_1 $,the relative anisotropic electric permittivity and magnetic permeability is proposed in  (12), the source $r_s < R_0$ and observer $r_o < R_0$, then the radial electromagnetic wave equal to the incident wave
\begin{equation}
E_r (\vec r_o ) = E_r^b (\vec r_o ),	
\end{equation}
\begin{equation}
H_r (\vec r_o ) = H_r^b (\vec r_o ),	
\end{equation}										
\hfill\break
${\boldsymbol{Theorem \ 6.5: }}$\ 
 Suppose that in the outer sphere annular layer of GLHUA double layer cloak, $R_1  < r \le R_2 $, the anisotropic relative electric permittivity and magnetic permeability parameter is proposed in (1); in the inner sphere annular layer GLHUA double cloak, $R_0  < r \le R_1 $, the relative anisotropic electric permittivity and magnetic permeability is proposed in  (12), the source $r_s < R_0$ and observer $r_o < R_0$, then the angular electromagnetic wave equal to the incident wave 
\begin{equation}	
\begin{array}{l}
 E_\theta  (\vec r_o ) = E_\theta ^b (\vec r_o ), \\ 
 E_\phi  (\vec r_o ) = E_\phi ^b (\vec r_o ), \\ 
 \end{array}
\end{equation}
\begin{equation}	
\begin{array}{l}
 H_\theta  (\vec r_o ) = H_\theta ^b (\vec r_o ), \\ 
 H_\phi  (\vec r_o ) = H_\phi ^b (\vec r_o ), \\ 
 \end{array}
\end{equation}	
\hfill\break
${\boldsymbol{Theorem \ 6.6: }}$\ 
 Suppose that in the outer sphere annular layer of GLHUA double layer cloak, $R_1  < r \le R_2 $, the anisotropic relative electric permittivity and magnetic permeability parameter is proposed in (1); in the inner sphere annular layer GLHUA double cloak, $R_0  < r \le R_1 $,the relative anisotropic electric permittivity and magnetic permeability is proposed in  (12), the source $r_s < R_0$ and observer $r_o < R_0$, then the electromagnetic wave equal to the incident wave
\begin{equation}
\vec E(\vec r_o ) = \vec E^b (\vec r_o ),
\end{equation}
\begin{equation}
\vec H(\vec r_o ) = \vec H^b (\vec r_o ),
\end{equation}

${\boldsymbol{Proof: }}$\ 
Summary of theorem 6.4 and theorem 6.5, we prove (141) and (142). Theorem 6.6 is proved. The theorem shown that the incident electromagnetic wave excited in the concealment $r_s < R_0$  can not be  disturbed by the cloak.

\section{GL electromagnetic Eikonal equation for anisotropic material in GLHUA cloak}
In previous sections, we present GL full wave field no scattering modeling and inversion for creating GHUA double layer cloak and proved the GLHUA cloak is invisible with concealment. The Geometry ray method can not be used for study our GLHUA cloak. The ray tracing is derived from Eikonal equation. The anisotropic material Eikonal equation is different from isotropic Eikonal equation.

\subsection {GL electromagnetic Eikonal equation for anisotropic material in spherical coordinate}

Using our GL Delta expansion in [3] and [4] for GL electromagnetic wave equation (19) and (23)  in [1], we obtain GL electromagnetic Eikonal equation and GL transport equation in GLHUA sphere and in GLHUA cloak. The GL electromagnetic Eikonal equation for anisotropic material in spherical coordinate is proposed in the following:

\begin{equation}
\begin{array}{l}
 \frac{1}{{\varepsilon _\theta  }}\left( {\frac{{\partial \psi _E }}{{\partial r}}} \right)^2  + \frac{1}{{\varepsilon _r }}\left( {\frac{1}{r}\frac{{\partial \psi _E }}{{\partial \theta }}} \right)^2  + \frac{1}{{\varepsilon _r }}\left( {\frac{1}{{r\sin \theta }}\frac{{\partial \psi _E }}{{\partial \phi }}} \right)^2  = \mu _\theta  , \\ 
 \frac{1}{{\mu _\theta  }}\left( {\frac{{\partial \psi _H }}{{\partial r}}} \right)^2  + \frac{1}{{\mu _r }}\left( {\frac{1}{r}\frac{{\partial \psi _H }}{{\partial \theta }}} \right)^2  + \frac{1}{{\mu _r }}\left( {\frac{1}{{r\sin \theta }}\frac{{\partial \psi _H }}{{\partial \phi }}} \right)^2  = \varepsilon _\theta  , \\ 
 \end{array}
\end{equation}
Where $T = \psi _E$ is the electric wave front, $T = \psi _H$ is the magnetic wave front, $\varepsilon _\theta   = \varepsilon _\phi  $ is the angular relative electric permittivity,
$\varepsilon _r$ is the radial relative electric permittivity, $\mu _\theta   = \mu _\phi  $
is the angular relative magnetic permeability, $\mu _r$  is the radial relative magnetic permeability. In section 3 ,in [1] we discovered and proved theorem 3.1, in the no source domain, with weight, $\sin \theta $,
sphere surface integral of radial GL electromagnetic wave is zero. It is the essential different between electromagnetic wave and acoustic wave. The additional constrain condition to be added in the Eikonal equation (143) and transport equation jointly to solve. 
\hfill\break
\subsection {GL electromagnetic Eikonal equation for anisotropic GHUA double layer cloak}

In the GLHUA double layer cloak,  $\varepsilon _r  = \mu _r  = 1$,  $\varepsilon _\theta   = \varepsilon _\phi   = \mu _\theta   = \mu _\phi$, in (1) for outer layer cloak, $R_1  < r \le R_2 $, and in (12) for inner layer cloak,$R_0  < r \le R_1 $,  the above two Eikonal equation in (143) become same equation. We proposed the Eikonal equation for GLHUA double layer cloak in the following
\begin{equation}
\frac{1}{{\varepsilon _\theta  }}\left( {\frac{{\partial \psi }}{{\partial r}}} \right)^2  + \left( {\frac{1}{r}\frac{{\partial \psi }}{{\partial \theta }}} \right)^2  + \left( {\frac{1}{{r\sin \theta }}\frac{{\partial \psi }}{{\partial \phi }}} \right)^2  = \varepsilon _\theta,  
\end{equation}
In the local no source domain of GLHHA outer layer cloak, with weight, $\sin \theta $, the sphere surface integral of radial GL electromagnetic wave is zero, also the radial GL electromagnetic wave is smoothly enter GLHUA outer layer cloak without scattering, when $r$ is going to boundary $r=R_1$, the GL electromagnetic wave is going to zero, these constrain equations to be adde in Eikonal equation (144) and transport equation jointly to solve.  Because $\mathop {\lim }\limits_{r \to R_1 } \varepsilon _\theta   = \mathop {\lim }\limits_{r \to R_1 } \mu _\theta   = \infty$, from the Eikonal equation (144), transport equation and above constrain equation, we obtained that wave front is discontinuous and splitting in spherical surface $r = R_1$. The ray propagation is complicated, some ray is terminated on boundary $r = R_1$ , some ray is born from $r = R_1$ [5]. It is shown that the wave phase velocity is less than light speed and tends to zero that making the electromagnetic wave propagation can not arrive to the boundary $r = R_1$. The traditional Eikonal equation and geometry ray method without intensity can not be used for study our GLHUA double cloak. The GL full wave field no scattering modeling and inversion is very important and powerful method for creating GLHUA cloak and proved that GLHUA cloak is invisible cloak with relative parameter not less than 1.

\section {Discussion and conclusion}
In this paper, we theoretically created GLHUA double layer cloak;
proved the relative parameers of the GLHUA double layer cloak are
not less than 1; the relative parameters and their derivative are
continuous across the outer boundary $ r R_2$ and inner boundary $r=R_0$;
We proved that the EM wave in the outside of the cloak can not be disturbed by the cloak; moreover we proved the EM wave field excited in the
outside of the cloak can not propagation penetract into the concealment.
We prove that, reciprocally, EM wave field excited in the concealment can not propagate to outside of the inner cloak. In the concealment of GLHUA double layer cloak, the EM wave field can not be disturbed by the cloak.
GL EM Eikonal equation for anisotropic material in GLHUA cloak
is proposed. From GL EM Eikonal equation, in GLHUA outer layer cloak,
 wave front is discontinuous and splitting; the ray propagation is complicated and discontinuous at boundary $r=R_1$.  
Summary, we theoretically proved the GLHUA double layer cloak
is invisible cloak with concealment. The relative electromagnetic parameters in
GLHUA double layer cloak are not less that 1 that makes that the the phase velocity of the electromagnetic
wave propagation in GLHUA double layer cloak is less than light speed and tends to zero at boundary $ r=R_1$; the reciprocal principle is satisfied in GLHUA double layer cloak. The GLHUA double layer cloak is practicable.
The practicalbe GLHUA double layer invisible cloak has important application in the space intersteller and spaceflight. Invisible science is for peace.
In December of 2016, we submitted 3 papers to arXiv for GLHUA double layer cloak with relative parameters not less than 1. The first paper arXiv:1612.02857 has already been published in arXiv. The second paper ¡°electromagnetic wave propagation in GLHUA sphere¡± has been submitted to arXiv with submission ID submit1762559. This is the third paper. The task and content
of the three paper are different from each other.
In my paper arXiv1706.10147[6], we find GLHUA exact analytic EM full wave solution of 3D MAXWELL equation in GLHUA double layer invisible cloak. 
GL simulation shows that there are novel GLHUA mirage bridge which is generated by the
GLHUA outer layer invisible cloak materials. 
The GLHUA mirage bridge make the GLHUA analytical EM wave propagation without exceeding light speed. The GLHUA mirage bridge  is obvious for $R_1=0.5R_2$ or $R_1=0.66R_2$,
for $R_> 0.8R_2$, the GLHUA mirage bridge wave become to the arc  GLHUA mirage wave on the inner spherical layer $r=R_1$, the obvious mirage bridge is disappear.  The analytical GLHUA EM wave show that in the any finite time the EM wave in GLHUA outer layer cloak can not be arrive to the inner spherical boundary,$r = R_1 $, that making the EM wave can not penetrate into the sphere $r < R_1 $. Therefore, the analytical GLHUA EM wave proved that GLHUA outer layer cloak is invisible cloak with sphere concealment $r < R_1 $. 
In generalized regularizing sense, when $r$ going to $R_1$, our GLHUA analytical EM wave field  are going to zero.

We publish this paper and paper [1] to arXiv are theoretical base and proof of our paper arXiv.org/abs/1612.02857.
Our three paper on GLHUA cloak and GLHUA spher and their theoretical proof to publish in arXiv are for open review.
Please colleague give comments to me by my email or give open comments in arXiv. 
copyright and patent of the GLHUA EM cloaks,GLHUA sphere and GL modeling
and inversion
methods are reserved by authors in GL Geophysical Laboratory.If some colleague
cite our papers in his work paper, please cite our paper as reference in his paper.

\begin{acknowledgments}
We wish to acknowledge the support of the GL Geophysical Laboratory and thank the GLGEO Laboratory to approve the paper
publication.
\end{acknowledgments}


\begin{thebibliography}{99}
\bibitem{2} 
Jianhua Li, F. Xie, Lee Xie, G.Xie, Novel GLHUA elctromagnetic invisible
double layer cloak with relative parameters not less than 1 and GL no scattering inversion, ArXiv 1612.02857
\end{thebibliography}
\end{document}